\documentclass[]{mn2e}
\usepackage{graphicx}
\usepackage{times}
\renewcommand{\vec}[1]{\bmath{#1}}

\title[Amplification and stability of magnetic fields and dynamo effect
in young A stars]{Amplification and stability of magnetic fields and dynamo effect
in young A stars}
\author[R. Arlt and G. R\"udiger]{R. Arlt\thanks{E-mail:
rarlt@aip.de} and G. R\"udiger
\\
Astrophysikalisches Institut Potsdam, An der Sternwarte 16, D-14482 Potsdam, Germany}
\begin{document}

\date{Accepted \dots. Received \dots; in original form \dots}

\pagerange{\pageref{firstpage}--\pageref{lastpage}} \pubyear{2010}

\maketitle

\label{firstpage}

\begin{abstract}
  This study is concerned with the early evolution of magnetic fields 
  and differential rotation of intermediate-mass stars which may 
  evolve into Ap stars. We report on simulations of the interplay of 
  differential rotation and magnetic fields, the stability limits
  and non-linear evolution of such configurations, and the prospects 
  of dynamo action from the unstable cases. The axisymmetric problem delivers
  a balance between field amplification and back-reaction of the
  magnetic field on the differential rotation. The non-axisymmetric
  case involves also the Tayler instability of the amplified toroidal 
  fields. We consider limits for field amplification and apply these 
  to young A stars. 

  Apart from its application to Ap stars, the instability is scrutinized for 
  the fundamental possibility of a dynamo. We are not looking for a dynamo as an 
  explanation for the Ap star phenomenon. The kinetic helicity is concentrated 
  near the tangent cylinder of the inner sphere of the computational domain
  and is negative in the northern hemisphere. This appears to be a
  ubiquitous effect not special to the Tayler instability. The latter
  is actually connected with a positive current helicity in the bulk
  of the spherical shell giving rise to a small, but non-vanishing $\alpha$-effect
  in non-linear evolution of the instability. 
\end{abstract}

\begin{keywords}
MHD -- stars: magnetic field -- stars: chemically peculiar.
\end{keywords}

\section{Introduction}
Stellar objects involve in many cases the amplification of
magnetic fields by differential rotation. The differential
rotation at the bottom of the solar convection zone, for example, acts
as a generator for toroidal magnetic fields while the differential
rotation is imposed from the convection zone. In principle,
there is a large domain of radiative zones in stars in which
the amplification by differential rotation is conceivable,
where no convective motions disturb the winding-up. This
concerns the whole range of intermediate-mass and massive 
stars from 1.2~solar masses up to several tens of a solar mass.
All these stars possess large radiative envelopes (apart 
from a possible, very thin convective surface layer). We will
often speak about A and Ap stars in this Paper, but the simulations
are actually applicable to a wider range of stars possessing
non-convective domains, such as solar-like stars with radiative
interiors or more massive B stars.

According to  St\c{e}pie\'n (2000), the pre-main-sequence
evolution of stars with masses near two solar masses is
accompanied by changes in angular momentum through disk
accretion, magnetic star-disk coupling, and magnetized
winds. All these effects likely lead to a differentially 
rotating interior. When approaching the zero-age main
sequence (ZAMS), the stars have established their radiative
envelopes, and accretion has ceased. It is then mainly by
magnetized winds that the angular velocity is reduced. This,
however, affects only the surface, while it takes extremely 
long for the small plasma viscosity to reduce also the angular 
velocity in the interiors. The coupling of the surface to the 
interior is strongly enhanced by magnetic fields. There are
a few cases for which an increase of the surface rotation 
period has actually been observed, such as V901 Ori 
(Mikula\v{s}ek et al. 2008).

Another source for differential rotation, especially for
stars with masses less than $2.5 {\rm M}_\odot$, may be the
convective phase in the pre-main-sequence evolution. While
there is no specific model available for a progenitor of
A stars, the computations by K\"uker \& R\"udiger (2008)
indicate tendencies of stronger differential rotation
for faster rotation, hotter surface temperatures, and
especially thinner convective envelopes. To which extent
the differential rotation will be present in the remaining
radiative star is debatable, since small magnetic fields
in the (then) radiative interior are sufficient to prevent 
differential rotation from spreading below the convection zone
(R\"udiger \& Kitchatinov 1997; Gough \& McIntyre 1998).

Studying the interaction between differential rotation and
magnetic fields is important for the understanding of the
existence of magnetic Ap stars, and for the explanation
why only a fraction of A stars show the peculiarities.
Differential rotation or magnetic fields or combinations
of the two may become unstable against small perturbations
(e.g.\ Watson 1981; Gilman \& Fox 1997; Dikpati et al.
2004; Braithwaite 2006b; Arlt et al. 2007; and many others).
Where MHD is a suitable description of the astrophysical
objects, we can group the instabilities into shear-driven,
current-driven, and buoyancy-driven instabilities.

Current-driven instabilities are particularly interesting
for toroidal magnetic fields, since these typically reach
the largest strengths if differential rotation is present.
Such toroidal fields also come along with currents giving
rise to a kink-type instability, whose most easily excited
modes are typically modes with low azimuthal wave number
(Vandakurov 1972; Tayler 1973; Spruit 1999). We refer to 
this instability here as Tayler instability.

Dynamo action arising from the Tayler instability was suggested
heuristically by Spruit (2002) and later in simulations 
by Braithwaite (2006a) who found sustained,
but not constant magnetic field energies in a differentially
rotating cylindric domain. Spherical, three-dimensional 
simulations of the solar radiative interior have 
been performed by Brun \& Zahn (2006) where the evolution
of a fossil field together with an imposed rotation profile
at the top of the domain (i.e. the bottom of the convection zone)
was investigated. The initial poloidal magnetic field as well as
the later combination of toroidal and poloidal fields showed
distinct instability signatures. The simulations were examined 
by Zahn, Brun \& Mathis (2007) on possible dynamo action, but none was 
found. The difference between the two attempts may be the presence 
of an enforced differential rotation inside the domain of Braithwaite's 
(2006a) simulations, as opposed to the enforcement on the boundary
only in the computations by Zahn et al. (2007). Another
non-linear simulation of the Tayler instability was performed
by Gellert, R\"udiger \& Elstner (2008) who also used a cylindrical setup
with a differential rotation enforced at the outer radial
boundary, but in this case depending on vertical direction
$z$ only. In terms of a dynamo-$\alpha$  they find values
of $|\alpha_{zz}| \sim 0.05$ measured against an imposed
vertical field. That $|\alpha_{zz}|$ is 100~per cent of the rms velocity
implying that the flow is entirely helical. However, all magnetic 
modes decay when the external magnetic field is switched off. 

The precise nature of the non-linear evolution of the Tayler instability
with a possible dynamo effect is thus a very interesting issue on
which we follow up here. The scenario of amplification of 
magnetic fields by differential rotation, the stability
of the resulting configurations, and the possible dynamo
effect in case of instability is very general to non-convective 
zones in a wide range of stars. Section~2 describes how the 
computations were set up, while Section~\ref{axisymmetric} shows
the results of the axisymmetric, non-linear simulations for the
field-amplification. The analysis in Section~\ref{linear_stability} 
deals with the stability of the results in Section~\ref{axisymmetric},
and Section~\ref{non-linear} evolves the full three-dimensional and
non-linear problem. Dynamo coefficients are derived in Section~\ref{dynamo}.
With respect to the instability, we will focus on the pre-main-sequence
evolution of Ap stars here, whereas the possible dynamo action is not
believed to be an important Ap star phenomenon, but is of general
interest for dynamo theory.

\section{Numerical setup}
The computational domain is a spherical shell with the radius $r$ 
running from an inner radius $r_{\rm i}$ to an outer radius $r_{\rm o}$ 
which is normalized to one by the stellar radius $R$ in the following.
The colatitude $\theta$ always covers the range from~0 to $\pi$; the
azimuthal coordinate $\phi$ runs from~0 to $2\pi$ in the non-axisymmetric
simulations. The inner radius
was set to $r_{\rm i}=0.5$ in all the computations of this study.

The investigation involves three steps: Step~I one is an 
axisymmetric simulation with an initial differential rotation $\Omega(s)$ 
where $s=r\sin\theta$ is the axis distance, and an initial poloidal 
magnetic field $\vec B = (B_r(r,\theta),B_\theta(r,\theta), 0)$.

Step~II is a stability analysis where a number of snapshots
of the simulation of the first step is used as background states
for a linear analysis of the stability of the system against
non-axisymmetric perturbations. We will restrict ourselves to
the stability of perturbations with azimuthal wave-numbers of $m=1$.

Step~III consists of fully non-linear, three-dimensional computations
of the evolution. In these simulations, the non-axisymmetric perturbation
is either present from the beginning, or is imposed at a later step as
an external `kick' to the system. We can thus compare the evolutions
of states which are either Tayler-stable or Tayler-unstable according
to Step~II. The non-linear evolutions also allow to assess the possibility
of driving a dynamo with the instability. Dynamo effect will be visible
only in terms of non-vanishing dynamo-coefficients of a mean-field
description, but not in growing and sustained magnetic fields, because
our system lacks a source of energy.

The computations employ the spherical MHD code by Hollerbach (2000)
which integrates the momentum, induction, and temperature equations in 
the Boussinesq approximation. We are not using temperature 
fluctuations in this study; the normalized, incompressible MHD 
equations are thus
\begin{eqnarray}
\frac{\partial{\vec u}}{\partial t} &=& 
  -(\vec u\cdot\nabla)\vec u + (\nabla\times \vec B)\times \vec B \nonumber\\
& &-\nabla p +{\rm Pm}\triangle\vec u,
\label{ns}\\
\frac{\partial{\vec B}}{\partial t} &=&
  \nabla\times  (\vec u\times \vec B) 
  +\triangle\vec B,
\label{induction}
\end{eqnarray}
where $\vec u$ and $\vec B$ are the velocity and magnetic fields and $p$ 
is the pressure. Additionally, the relations $\nabla\cdot\vec u=0$ and 
$\nabla\cdot\vec b=0$ hold. The magnetic permeability and the density are set 
to unity in this system of units. Since all quantities are normalized with
the magnetic diffusivity $\eta$, the magnetic Prandtl number ${\rm Pm}=\nu/\eta$
appears in the momentum equation, where $\nu$ is the kinematic viscosity.
Velocities are normalized by $\eta/R$, times by $R^2/\eta$.

For Steps~I and III, the differential rotation and the poloidal magnetic field are 
initial conditions. They evolve freely without any further 
imposed properties. The profile of the initial angular velocity follows
\begin{equation}
  \Omega(s) = \frac{\Omega_0}{\sqrt{1+s^{\,q}}},
  \label{omega}
\end{equation}
where $\Omega_0$ is the non-dimensional angular velocity on the axis and $q$ is
a parameter controlling the steepness of the rotation profile.
With our normalization, the initial magnetic Reynolds number
is actually ${\rm Rm}\equiv\Omega_0$, since
\begin{equation}
  {\rm Rm} = \frac{R^2 \Omega_\ast}{\eta},
\end{equation}
where $\Omega_\ast$ is the angular velocity of the star in
physical units. The magnetic Reynolds number was ${\rm Rm} =
20\,000$ in all computations except one; the magnetic Prandtl number was
always ${\rm Pm} = 1$. 

The pre-main-sequence evolution of the rotation period of 
intermediate-mass stars contains most likely both a spin-up 
(contraction and accretion) phase and a spin-down phase 
with magnetized winds (St\c{e}pie\'n 2000).
We are concerned with the evolution near the end of the
pre-main-sequence phase and the early main-sequence life
when most of the star is already radiative (convectively
stable). The probable angular-momentum change is a spin-down then,
acting on the surface of the star, and this is the reason why we 
assumed an angular velocity {\it decreasing\/} with axis distance.

The initial profile uses $q(t=0)=4$, but $q$ will later be used to 
measure the actual steepness of the rotation curve in the 
simulation. The specific angular momentum $s^2\Omega(s)$ is
increasing with axis distance everywhere in the computational
domain. The system is thus not prone to the hydrodynamic Rayleigh 
instability which we do not want to investigate here. The hydrodynamic
stability has been tested numerically with non-axisymmetric 
perturbations which were symmetric and anti-symmetric with respect
to the equator. All situations delivered a decay of the perturbations.

Magnetic fields are measured in terms of Lundquist numbers, which
is the same as the non-dimensional Alfv\'en velocity in our system
of units,
\begin{equation}
  B = \frac{R B_{\rm phys}}{\sqrt{\mu\rho}\, \eta}.
  \label{bconvert}
\end{equation}
The magnetic diffusivity $\eta$ in time-dependent simulations
typically represents a value between the microscopic diffusivity
of the plasma and the turbulent diffusivity resulting from, e.g.,
averaged convective motions. Of the quantities entering (\ref{bconvert}),
$\eta$ is the one which is known least. It is therefore best to eliminate $\eta$ 
by ${\rm Rm}$ and thus retrieve the physical magnetic fields by comparing 
its Alfv\'en speed with the rotational velocity,
\begin{equation}
  B_{\rm phys} = \sqrt{\mu\rho} \Omega_{\rm phys} R \frac{B}{\rm Rm}
  \label{bphys}
\end{equation}
Considering a $2M_\odot$ star with a radius of $R_\ast = 1.5R_\odot$,
a rotation period of 10~days ($\Omega_{\rm phys}=7.3\cdot10^{-6}$)
and a density of 0.015~g/cm$^3$ at
$0.75R_\ast$, a magnetic field of $B=100$ converts to 1660~G
in physical units with the magnetic Reynolds number of 20\,000 used in
most of our simulations.

The boundary conditions for all runs presented here employ
stress-free conditions for the flow and vacuum conditions
for the magnetic field at both the inner and outer radii,
$r_{\rm i}$ and $r_{\rm o}$. Although the inner sphere is
supposed to be highly conducting, a vacuum condition helps
in obtaining smoother solutions compared to a superconducting
condition. The latter prohibits a penetration of the magnetic field
through the inner boundary. This typically leads to strong currents
near the inner boundary which are not there if the conductivity 
would change continuously into the inner sphere. If, for 
efficiency reasons, the whole star cannot be computed, the vacuum
condition is a suitable choice despite the non-vanishing conductivity.
We will come back to a test run with a perfect-conductor
boundary at $r_{\rm i}$ in Section~\ref{non-linear}.

Note that the assumption
of an initially independent angular velocity of the vertical
axis $z=r\cos\theta$ (Tayler-Proudman state) is not compatible 
with a stress-free condition on a spherical surface. This
causes small meridional flows and results in a small deviation 
from the Tayler-Proudman state. However, this is a much better
choice than an initial rotation profile depending on $r$ which 
causes a severe redistribution of angular momentum towards a configuration 
which is very near a Tayler-Proudman state during the first rotations
of the system. As we did not want to obtain a mixed evolution
of this hydrodynamic phenomenon with the magnetic phenomena,
we chose $\Omega(s)$ instead of more complicated profiles
$\Omega(r,\theta)$. 

The spin-down by winds during the late pre-main-sequence
phase of intermediate-mass stars is probably not causing
a precise $\Omega(s)$ profile, but for the sake of
physical clarity, we choose a Tayler-Proudman state as an
initial condition for the runs. Note that the removal of
angular momentum by a magnetized wind is not implemented
in our setup, so the simulations can be understood as
mimicking a period around the time when the star enters the
main sequence.

\begin{figure}
\centering
\includegraphics[width=0.48\textwidth]{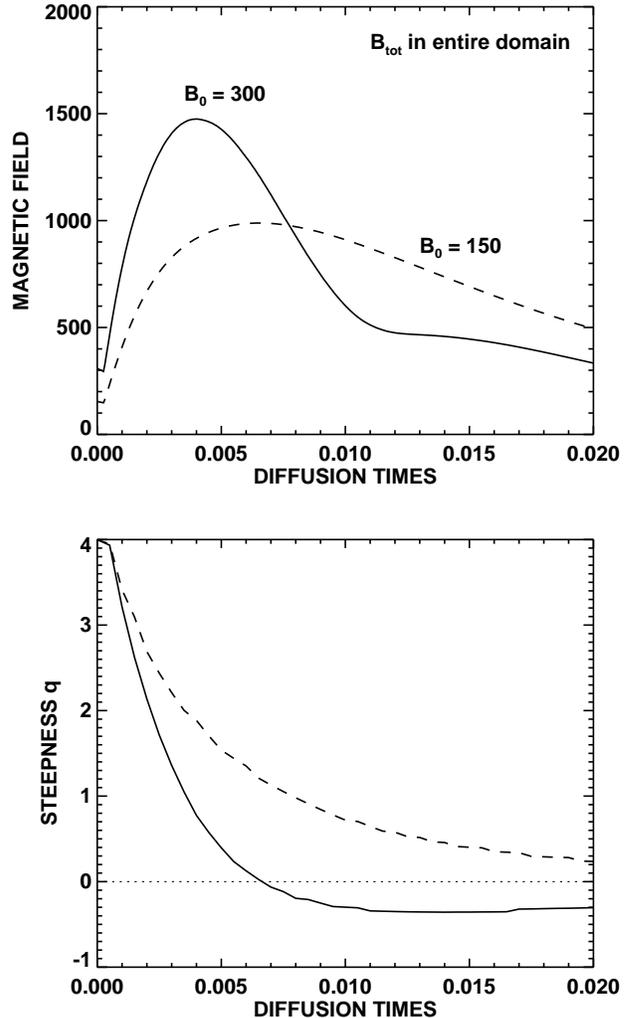}
\caption{Amplification of the toroidal magnetic field under
the influence of differential rotation, starting from a poloidal field. 
The solid line refers to an initial maximum magnetic field strength
of 300, the dotted line is the evolution of an initial
field strength of 150.
\label{amplification}}
\end{figure}

\section{Axisymmetric evolution (step I)}\label{axisymmetric}
First we study the amplification of magnetic fields by shear 
as well as the back-reaction of the fields on the differential
rotation. The two-dimensional, non-linear simulations thus start 
with an initial differential rotation and a purely poloidal 
magnetic field in a radiative stellar zone. These initial 
conditions evolve freely without any imposed flows or fields,
neither in the bulk of the computational domain nor at the
boundaries.

The early phase of the simulation shows a generation and steep 
amplification of toroidal magnetic field through differential 
rotation. The generated $B_rB_\phi$ and, to a smaller extent, 
also the $B_\theta B_\phi$ impose a Lorentz force to the rotational
velocity and redistributes angular momentum. This is why at the
same time of field amplification, the differential rotation 
starts to decrease, and the toroidal-field growth is thus limited.
The whole process reaches a maximum of magnetic
field energy after a few rotational periods. 

Fig.~\ref{amplification} shows the maximum magnetic field 
in the spherical shell as a function of time. The magnetic 
Reynolds number was ${\rm Rm} = 20\,000$ and the magnetic 
Prandtl number was ${\rm Pm}=1$. Two runs with initial,
purely poloidal magnetic fields with maximum strengths of 
$B_0=300$ and $B_0=150$ are shown.
The lower panel of the Figure shows the decay of the differential
rotation. We assume that the differential rotation roughly follows a
profile (\ref{omega}) throughout the simulation, with varying 
parameters $\Omega_0$ and $q$. The values of $\Omega(s)$ in the
equatorial plane of each snapshot of the simulation are used to
fit a best profile (\ref{omega}) delivering two time series for
$\Omega_0(t)$ and $q(t)$. It is the temporal
evolution of $q$ which is shown in the lower panel of
Fig.~\ref{amplification}. Interestingly, the run with a 
stronger initial poloidal magnetic field shows a reversal of
the differential rotation profile after 0.006 diffusion times.
Only after a very long period, not plotted here, the shear
converges to zero in the entire computational domain.

Maximum internal field strengths are about 24~kG in the run
with $B_0 = 300$ and about 16~kG in the run with $B_0 = 100$,
using the conversion (\ref{bphys}).
Much higher magnetic Reynolds numbers encountered in real stars
lead to stronger toroidal fields. A run with ${\rm Rm}=40\,000$
later described in Section~\ref{non-linear} reached about 46 kG.

The amplification time results from the comparison of the Lorentz
force acting on the differential rotation with the generation of
toroidal magnetic field by the shear. The linearized action of the
Lorentz force may be approximated by $\Delta \Omega \sim \Delta t 
B_r B_\phi/\mu\rho R$ and the linear winding up of toroidal field 
in the induction equation can be estimated by $B_\phi \sim \Delta t 
\Delta\Omega B_r / R$. Plugging the $B_\phi$ into the first
relation, delivers $\Delta t = R\sqrt{\mu\rho}/B_r$. The amplification
time is independent of the rotation rate, the amplitude of the
differential rotation, and the diffusivity. In our dimensionless
system, the amplification time is thus $\Delta t=0.0033$ for the
initial field of $B_0=300$. This is very close the the peak
time of the solid line in Fig.~\ref{amplification}. The dashed
curve from half the initial field strength peaks at roughly
twice that period -- a bit earlier since diffusion and hydrodynamic
redistribution of angular momentum are also at play in this
numerical setup. There is a slight tendency of smaller magnetic
Reynolds numbers producing shorter times to reach the peak
magnetic field. This is solely due to the non-vanishing viscosity,
being present in our system of finite ${\rm Rm}$, which adds to the reduction 
of the differential rotation and limits the growth eventually.

A real system with stellar microscopic diffusivities of say 
$\eta=1000$~cm$^2$/s will show a nearly stationary phase 
after the amplification, since the Ohmic decay times are of 
order gigayears. A time-dependent numerical simulation cannot model 
the very low diffusivity at the true rotation rate, and the solution 
is subject to a much larger Ohmic dissipation after the maximum toroidal 
field is reached. The following process is thus characterized by a decay 
of both magnetic fields and differential rotation. Since the simulations 
preserve angular momentum parallel to the rotation axis, the asymptotic 
state is field-free and with uniform rotation. We are not interested 
in this asymptotic behaviour.
Intermediate-mass stars are not living long enough to show
this final state during their evolution on the main sequence.

\begin{figure}
\centering
\includegraphics[width=0.48\textwidth]{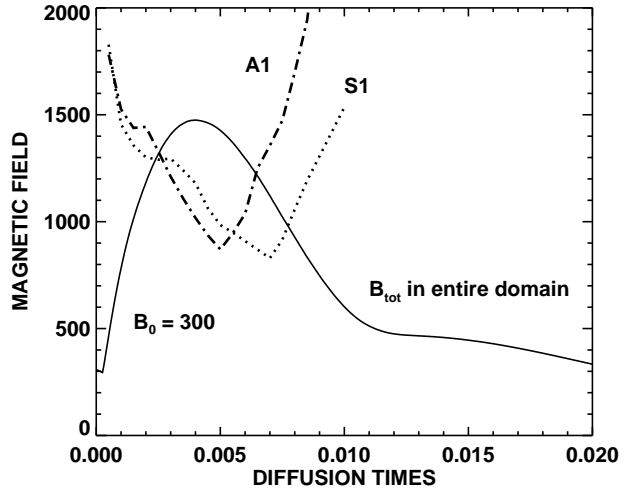}
\caption{Stability limits for the magnetic fields and velocity 
fields of each snapshot of the axisymmetric evolution of the 
$B_0=300$ case. The solid line is copied from Fig.~\ref{amplification}.
The stability limit is plotted for a perturbation of $m=1$, the dotted
line refers to a perturbation whose velocity field is symmetric 
with respect to the equator, the dash-dotted line shows the 
antisymmetric perturbation.
\label{stability}}
\end{figure}

\section{non-axisymmetric, linear stability (step II)}\label{linear_stability}
Differential rotation and rotation in general stabilize 
any current-driven instabilities (Tayler instability) as 
was shown by R\"udiger \& Kitchatinov (2010). Differential 
rotation together with diffusion has a destructive effect 
on any non-axisymmetric structures. At some point during 
the amplification of magnetic fields by shear, the 
axisymmetric solution of Section~\ref{axisymmetric} may become 
unstable against non-axisymmetric perturbations, as soon as
the stabilizing effect of the differential rotation ceases.

Each snapshot of the run with $B_0=300$ from Section~\ref{axisymmetric}
was thus tested individually on whether or not a non-axisymmetric 
perturbation can grow on the particular axisymmetric configuration. 
However, from the snapshots, only the azimuthal velocity and the 
toroidal magnetic field, $u_\phi$ and $B_\phi$, are used for the 
stability analysis. The full problem was found to be numerically 
demanding and is postponed to a future study. The final conclusions 
are not affected by this simplification.

We consider the linearized 
equations for a given azimuthal wave number $m$. The snapshots 
delivered background states in $\vec u$ and $\vec B$ which 
affect the perturbation $(\vec u', \vec B')$ and lead to 
exponential growth or decay of the perturbation.

The stability analysis is based on the linear, normalized MHD
equations
\begin{eqnarray}
\frac{\partial{\vec u'}}{\partial t} &=& \bigl[\,
  \vec u'\times\nabla\times \vec u + \vec u\times\nabla\times \vec u'
  -\nabla(\vec u'\cdot \vec{u})\,\bigr] \nonumber\\
& &+S\bigl[\,(\nabla\times \vec B')\times \vec B + (\nabla\times \vec B) \times \vec B'\,\bigr] \nonumber\\
& &-\nabla p +{\rm Pm}\triangle\vec u',
\label{ns_lin}\\
\frac{\partial{\vec B'}}{\partial t} &=&
  \nabla\times(\vec u\times \vec B' + S\, \vec u'\times \vec B )
  -\triangle\vec B'\,,
\label{induction_lin}
\end{eqnarray}
again with $\nabla\cdot\vec u'=0$ and $\nabla\cdot\vec B'=0$. The 
constant density $\rho$ and the magnetic permeability $\mu$ are also 
unity in this normalization.$S$ is a factor which scales the total 
background magnetic field taken from the axisymmetric run. The critical 
$S$ for instability is determined. The background is unstable, if $S>1$.
Note that we do not evolve the system actually in time: the axisymmetric 
background velocity and magnetic fields are constant. The time-dependence 
appears here solely for the test on growing or decaying perturbations.

The dotted and dot-dashed lines in Fig.~\ref{stability} are the results
of the stability analysis. The evolution derived from the axisymmetric 
simulation starting with an initial maximum field strength of 300 is 
taken from Fig.~\ref{amplification} and is plotted as a solid line. By
S1 we refer to a velocity perturbation $\vec u'$ which is symmetric 
with respect to the equator and has $m=1$, A1 is the corresponding 
antisymmetric perturbation. If at any given time both stability lines 
are above the solid line, the corresponding snapshot is stable against 
$m=1$ perturbations. The first snapshots with relatively weak toroidal 
fields are all stable. The stability lines cross the solid one at about 
$t=0.0023$. There is a minimum marginal stability at $t=0.0055$ for the 
S1 mode, and at $t=0.0070$ for the A1 mode.

After these times, the stability limits increase again. The stabilization
may be due to the change of sign in the shear. The differential rotation
is then non-vanishing again and may impose a stabilizing effect on
perturbations despite its relatively small, positive amplitude.

Higher $m$ require similar magnetic fields for instability as was shown
in test runs with $m=2$.

\begin{table}
\caption{Four non-linear simulations with their parameters as 
discussed in Sections~\ref{non-linear} and~\ref{dynamo}. The truncations
for the Chebyshev, Legendre, and Fourier modes are given
as $K$, $L$, and $M$, respectively. The column $t_{\rm pert}$ gives
the time when the non-axisymmetric perturbation was injected into the
system, expressed in diffusion times.}
\label{runs}
\begin{tabular}{lcccl}
\hline
Run     & ${\rm Rm}$ & ${\rm Pm}$ & $K \times L\times M$    &  $t_{\rm pert}$ \\
\hline
NL000   & 20000 & 1  & $35 \times 80\times 80$ &  0 \\
NL003   & 20000 & 1  & $40 \times 60\times 60$ &  0.003 \\
NL003h  & 40000 & 1  & $40 \times 60\times 60$ &  0.003 \\
NL005   & 20000 & 1  & $40 \times 60\times 60$ &  0.005 \\
\hline
\end{tabular}
\end{table}

\begin{figure}
\centering
\includegraphics[width=0.48\textwidth]{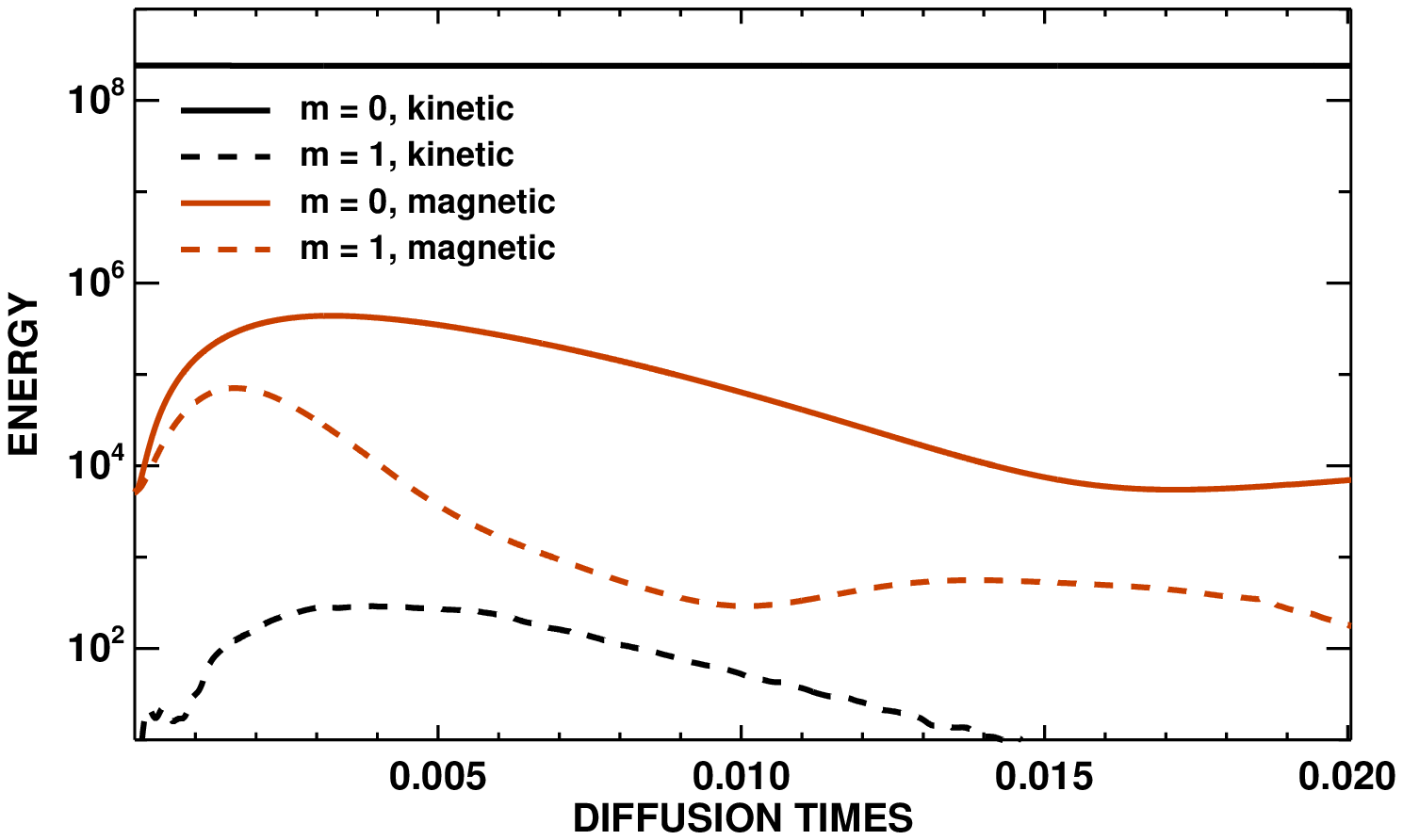}
\caption{Evolution of a non-axisymmetric perturbation added 
to the axi\-symmetric run of Fig.~\ref{amplification} at $t=0$.
\label{plotfile_nonlin11c}}
\end{figure}

\begin{figure}
\centering
\includegraphics[width=0.48\textwidth]{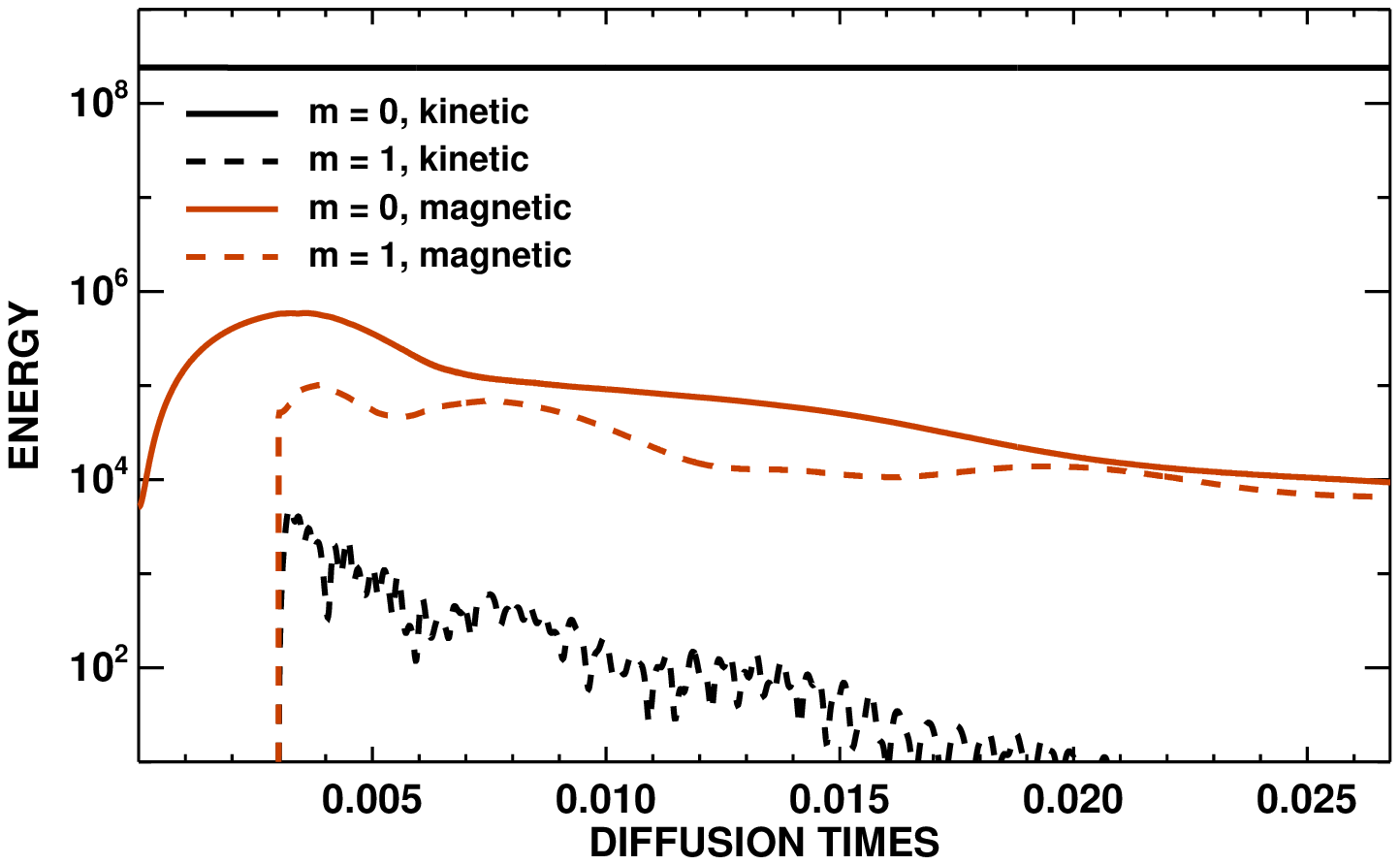}
\caption{Evolution of a non-axisymmetric perturbation added 
to the axi\-symmetric run of Fig.~\ref{amplification} at $t=0.003$.
\label{plotfile_nonlin16a}}
\end{figure}

\begin{figure}
\centering
\includegraphics[width=0.48\textwidth]{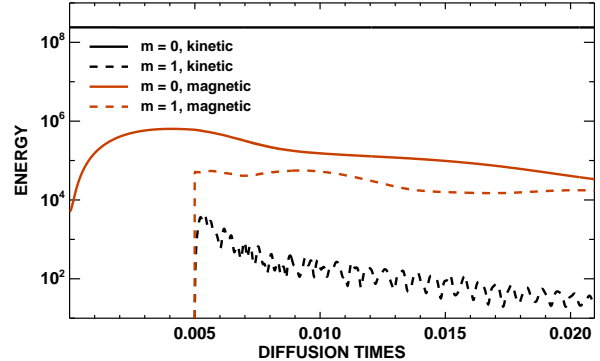}
\caption{Evolution of a non-axisymmetric perturbation added 
to the axi\-symmetric run of Fig.~\ref{amplification} at $t=0.005$.
\label{plotfile_nonlin15d}}
\end{figure}

\section{non-linear evolution in 3D (step III)}\label{non-linear}
The initial conditions are the same as for Section~\ref{axisymmetric},
but the system is now extended to include also a Fourier decomposition
of $\vec u$ and $\vec B$ in azimuthal direction. The main 
difference is a non-axisymmetric perturbation hitting the system at a 
given time $t_0$. The perturbation is applied to $\vec B$, has an 
azimuthal wave number of $m=1$ and is symmetric with respect to the 
equator, i.e. $B_r'(\theta) = B_r'(\pi-\theta)$, 
$B_\theta'(\theta) = -B_\theta'(\pi-\theta)$, and 
$B_\phi'(\theta) = B_\phi'(\pi-\theta)$. The resulting flow is thus
antisymmetric and can be compared with the A1-mode in the linear
stability diagram in Fig.~\ref{stability}. A small overview of the
models is given in Table~\ref{runs}.

Figs.~\ref{plotfile_nonlin11c}--\ref{plotfile_nonlin15d} show
the evolution of the energies of the lowest azimuthal modes,
$m=0$ and $m=1$, for three different instances of perturbations,
$t=0$, $t=0.003$, and $t=0.005$. The main feature of the run 
with the perturbation at the beginning is the quicker decay of the 
energy in the non-axisymmetric magnetic field as compared with the
other two simulations. Their perturbations at $t=0.003$, 
and $t=0.005$ are within the unstable window (cf.~Fig.~\ref{stability}) 
and show a much more persistent presence in the system. The energy of the 
magnetic $m=1$ mode may even become nearly as large as the magnetic $m=0$ 
energy. It is interesting to note that the energy in the $m=1$ 
mode of the velocity field decays more rapidly than the magnetic 
energy in both unstable cases.

\begin{figure*}
\centering
\includegraphics[width=15cm]{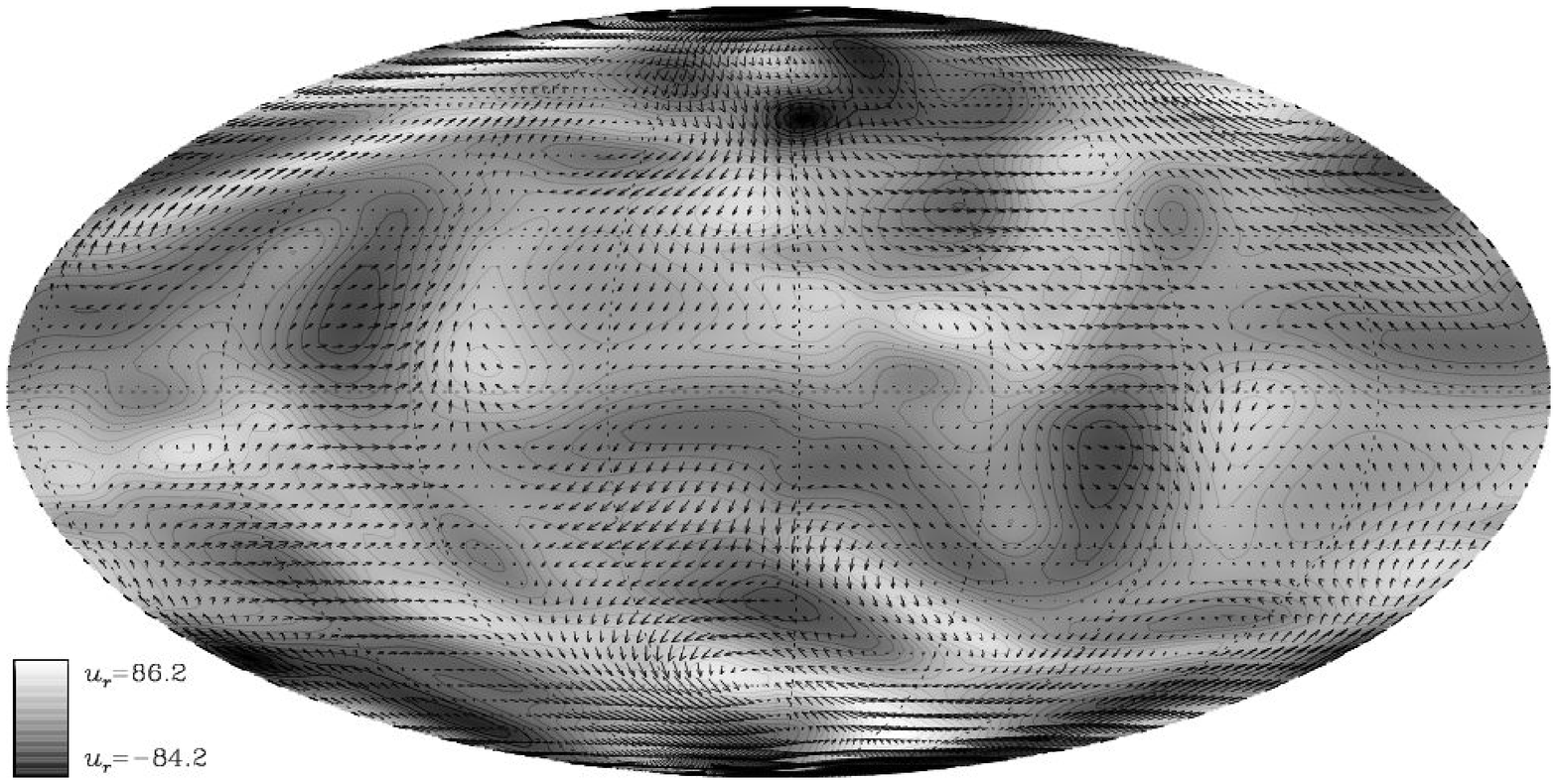}
\includegraphics[width=15cm]{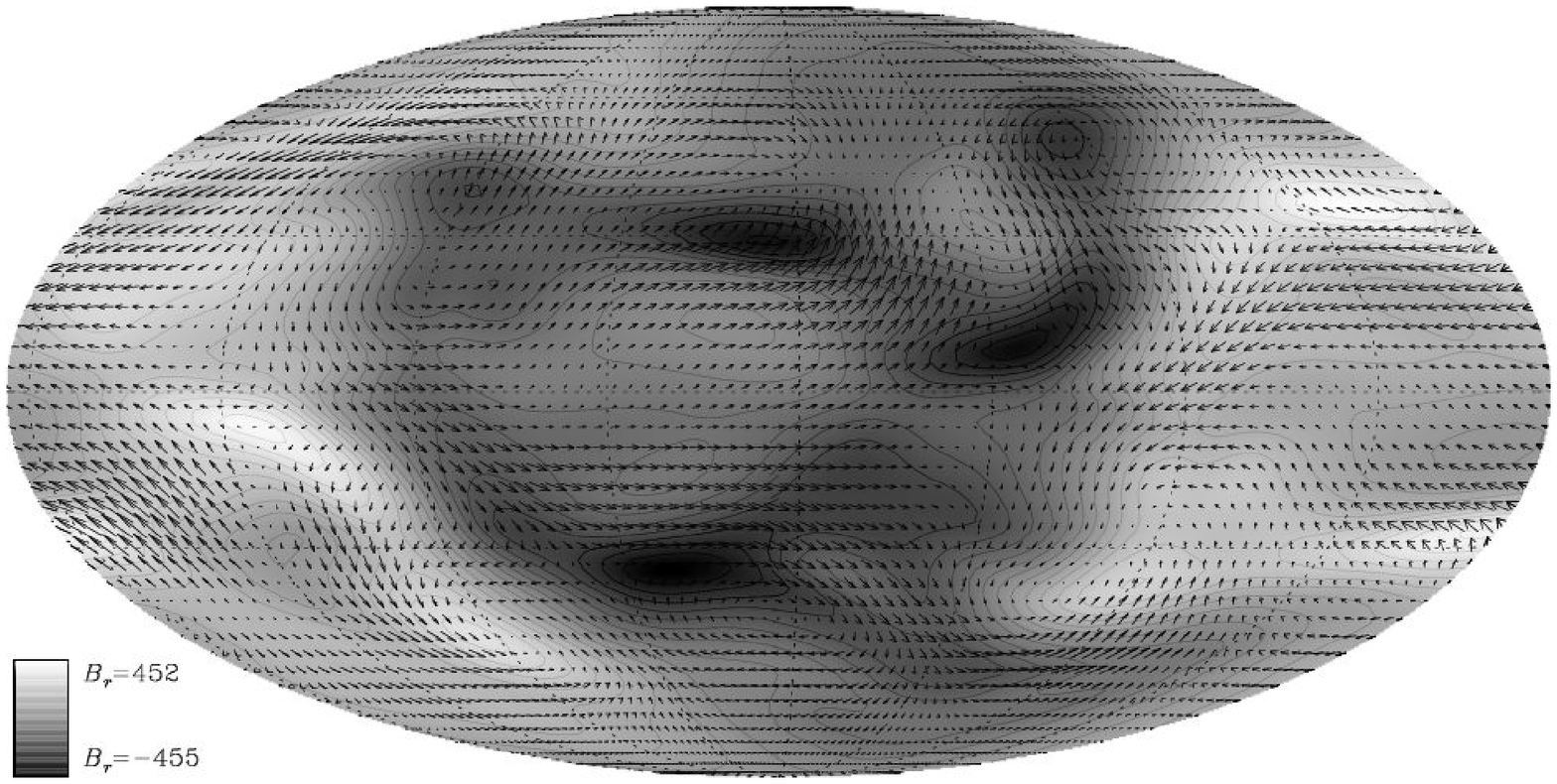}
\caption{non-axisymmetric velocity field (top) and magnetic field (bottom) 
at $r=0.75$ and $t=0.005$ from run NL003. The axisymmetric parts (fast rotation 
and strong toroidal magnetic field) were subtracted from the snapshot before 
plotting. The contours represent the radial components, while the arrows 
are the projected field directions on the $(\theta,\phi)$-surface. Note that
the actual resolution for the computation of the non-linear terms in both 
horizontal directions, $\theta$ and $\phi$, is twice as high as plotted here.
\label{spsurfm_nonlin16a}}
\end{figure*}

\begin{figure}
\centering
\includegraphics[width=0.48\textwidth]{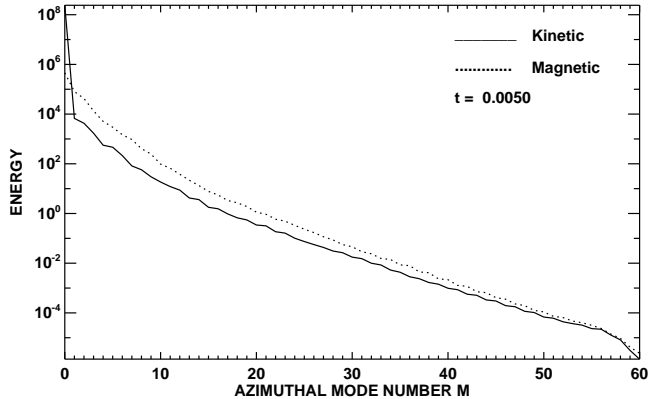}
\caption{Distribution of kinetic energy (solid line) and magnetic 
energy (dotted line) over the azimuthal wave number $m$ for the
run NL003 at $t=0.005$, i.e. 0.002~diffusion times after the 
perturbation was injected.
\label{spene_nonlin16a_250}}
\end{figure}

Typical horizontal slices of the velocity and magnetic fields
are shown in Fig.~\ref{spsurfm_nonlin16a}. They are plotted
for $r=0.75$ which is half-way between the inner and outer radial 
boundaries. Azimuthal mode numbers $m>1$ are obviously at play, but 
we cannot talk about a turbulent state here. The $m=2$ contains roughly
half the magnetic energy of the $m=1$ mode throughout the simulation. 
An example spectrum for the kinetic and magnetic energies of
all individual $m$-modes is shown in Fig.~\ref{spene_nonlin16a_250}. 
The energy in the $m=4$ mode is already two orders of magnitude 
smaller than the energy in $m=1$. The total contrast between $m=1$ and
$m=60$ is $10^{10}$ in the kinetic energy and one order of magnitude 
higher in the magnetic energy. Strongest magnetic fields appear to be 
concentrated in low latitudes. Since the axisymmetric parts of $\vec u$ 
and $\vec B$ were subtracted before plotting Fig.~\ref{spsurfm_nonlin16a}, 
the surface maps also reflect the energy ratio of magnetic $m=1$
energy to kinetic $m=1$ energy. While the ratio of $|B_r|$ to $|u_r|$ 
is 5.3, the energy ratio would be almost 30, being compatible with the 
$m=1$ energy ratio in Fig.~\ref{plotfile_nonlin16a} between $t=0.003$ and 
$t=0.005$. The magnetic fluctuations are dominating over the velocity
fluctuations in these simulations.

\begin{figure}
\centering
\includegraphics[width=0.45\textwidth]{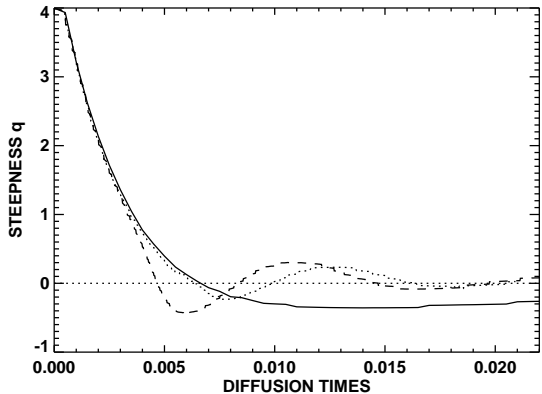}
\caption{Variation of the steepness of the rotation profile
with time: the solid line is the result from the axisymmetric
run and is the same curve as in Fig.~\ref{amplification}; the
dashed line is the result from the non-linear run with a 
non-axisymmetric perturbation at $t=0.003=9.5 P_{\rm rot}$, and 
the dotted line is for a perturbation at $t=0.005=15.9 P_{\rm rot}$.
\label{omega_comparison}}
\end{figure}

The impact of the non-axisymmetric instability on the rotation
profile is shown in Fig.~\ref{omega_comparison}. An enhanced 
angular momentum transport to reduce the differential rotation
is only notable for the simulation with a perturbation at
$t=0.003$. The most obvious difference between the unperturbed
run and the unstable runs is the time it takes to reach fully 
uniform rotation. Both perturbed simulations reach a state of
uniform rotation after 0.017 diffusion times, while it takes
the axisymmetric run about 0.05 diffusion times to be similarly
uniform in rotation (outside the plotting window).

\section{Dynamo effect}\label{dynamo}
The following section is not directly related to the pre-main-sequence
evolution of Ap stars. We are interested in a possible dynamo 
effect arising from the instability, since this will open very
interesting possibilities of magnetic-field generation in various
contexts. We are not proposing the dynamo effect to be overly
important for Ap stars. 

The non-linear interaction of the unstable 
$m=1$ mode may generate an axisymmetric amplification effect of 
poloidal fields out of toroidal fields. In that case, the Tayler 
instability would provide a closed dynamo loop besides the 
generation of toroidal fields by differential rotation. Sustained
dynamo action is conceivable as long as there is a mechanism
that sustains differential rotation, whence toroidal fields prone
to instability. Rotating convection zones are such providers of 
differential rotation, for example.

It is assumed here that the dynamo action can be described by a 
non-vanishing turbulent electromotive force (EMF) delivering a large-scale
axisymmetric magnetic field from the non-axisymmetric instability 
state (which is not necessarily a turbulent state). Everything axisymmetric
is considered large-scale, all non-axisymmetric contributions are 
considered small-scale in this context. We are also
assuming that the mean-field coefficients only relate the large-scale
magnetic field $\overline{\vec B}$ and its first derivatives with
the EMF, and use the decomposition
\begin{eqnarray}
  {\rm\bf EMF}&=&\phantom{-}\vec\alpha \overline{\vec B} + \vec\gamma \times \overline{\vec B}
  -\vec\beta(\nabla\times\overline{\vec B}) \nonumber\\
  &&- \vec\delta\times (\nabla\times\overline{\vec B})
  -\vec\kappa(\nabla\overline{\vec B})^{(\rm sym)}
\label{emf}
\end{eqnarray}
into symmetric $\vec\alpha$- and $\vec\beta$-tensors, the vectors 
$\vec\gamma$ and $\vec\delta$, and the third-rank tensor
$\vec\kappa$ acting on the symmetric part of the tensor gradient
of $\overline{\vec B}$.

When restricting to the diagonal elements of $\vec\alpha$, one
notes that only $\alpha_{\phi\phi}$ can cause an axisymmetric
growth of the large-scale poloidal magnetic field from an axisymmetric
large-scale toroidal field. Remember that the non-axisymmetric Tayler 
instability is now supposed to be hidden in the mean-field coefficients,
and is not explicitely visible in this description. As long as we 
are looking for the driving of an axisymmetric dynamo, the 
existence of the Tayler instability and its properties are supposed
to be entirely comprised by the mean-field coefficients.

\subsection{Existence of an $\alpha_{\phi\phi}$}
Our first attempt to look for dynamo action thus consists of
searching for a correlation between the $\phi$-component of the turbulent 
electromotive force, ${\rm tEMF} = \langle\vec u'\times \vec B'\rangle_\phi$, 
and the $\phi$-component of the large-scale magnetic field, $\langle B_\phi\rangle$.
The brackets refer to suitable averages in space 
and, ideally, also in time. We simplify the analysis further
by using the azimuthal direction for the spatial average and discard
time averages. This type of averaging will be denoted by 
overbars, hence ${\rm tEMF}_\phi = \overline{(\vec u'\times \vec B')}_\phi$ and 
$\overline{B_\phi}$. The angular brackets will be used for averages over entire
hemispheres of the computational domain in the following.

Instead of averaging over time, we rather look at the temporal evolution
of the correlation between ${\rm tEMF}_\phi$ and $B_\phi$. The fluctuating 
quantities are thus derived by $\vec u'=\vec u-\overline{\vec u}$ and 
$\vec B'=\vec B-\overline{\vec B}$, where overbars
are always $\phi$-averages. The values of the average turbulent
EMF and $\overline{B_\phi}$ deliver scatter plots from all the 
radial and latitudinal grid points of an entire hemisphere.
The slope of the regression line as to represent $\alpha_{\phi\phi}$
and the correlation coefficient were derived from these sets
of pairs $\{{\rm tEMF}_\phi(r,\theta),\overline{B_\phi}(r,\theta)\}$
from each snapshot in time and each individual simulation.


We also computed the kinetic helicity in the two hemispheres
(N and S) from the velocity fluctuations by
\begin{equation}
  {\cal H}_{\rm N} = \left\langle \vec u'\cdot {\rm curl} \vec u'\right\rangle_{\rm N},
  \qquad
  {\cal H}_{\rm S} = \left\langle \vec u'\cdot {\rm curl} \vec u'\right\rangle_{\rm S},
\end{equation}
where averages are taken over entire hemispheres. It will be 
interesting to see to which extent the kinetic helicity 
is related to what we measure as an $\alpha$-effect. Turbulence 
driven by convection leads to an 
\begin{equation}
 \alpha = -\frac{1}{3}\tau_{\rm cor}\langle\vec u'\cdot {\rm curl} \vec u'\rangle,
 \label{helicity}
\end{equation}
where $\tau_{\rm cor}$ is the correlation time.
This has also been shown in various simulations (e.g. Giesecke, Ziegler \& R\"udiger 2005; 
K\"apyl\"a, Korpi \& Brandenburg 2009).
Finally, the root mean square (rms) values of the velocity fluctuations
are computed for the entire computational domain for the same
snapshots in time.

We will again refer to the three simulations with perturbations
at $t=0$, $t=0.003$, and $t=0.005$. We recall that in the first one,
the perturbation was added to a linearly stable state while in the
other simulations, the perturbations were added to linearly unstable
cases. In the last case, the toroidal field was already past its
maximum value of $\overline{B_\phi}=1434$, but had still a supercritical strength 
of $\overline{B_\phi}=1361$ (we recall that the maximum initial poloidal field 
strength was $B_0=300$ for comparison). 

\begin{figure}
\includegraphics[width=0.44\textwidth]{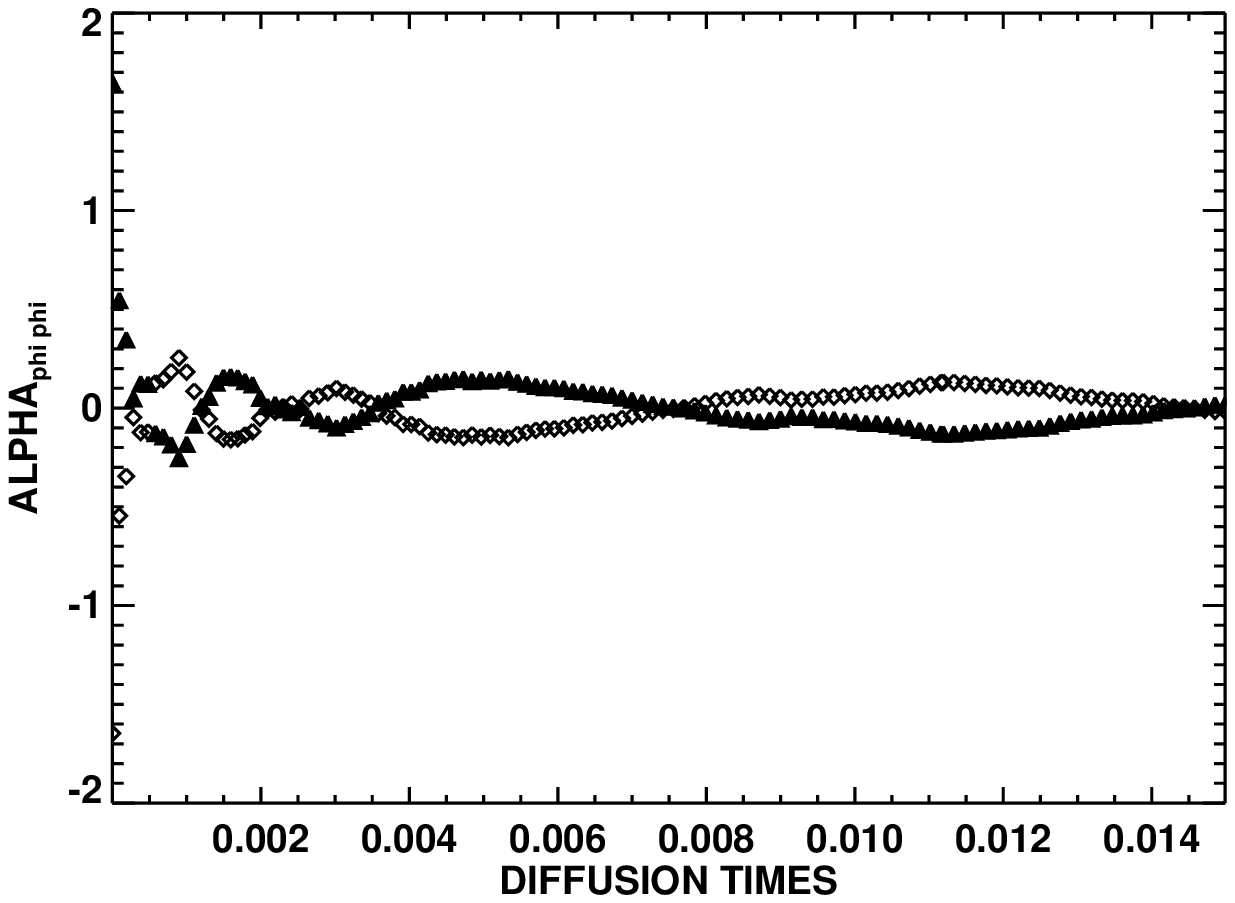}
\includegraphics[width=0.44\textwidth]{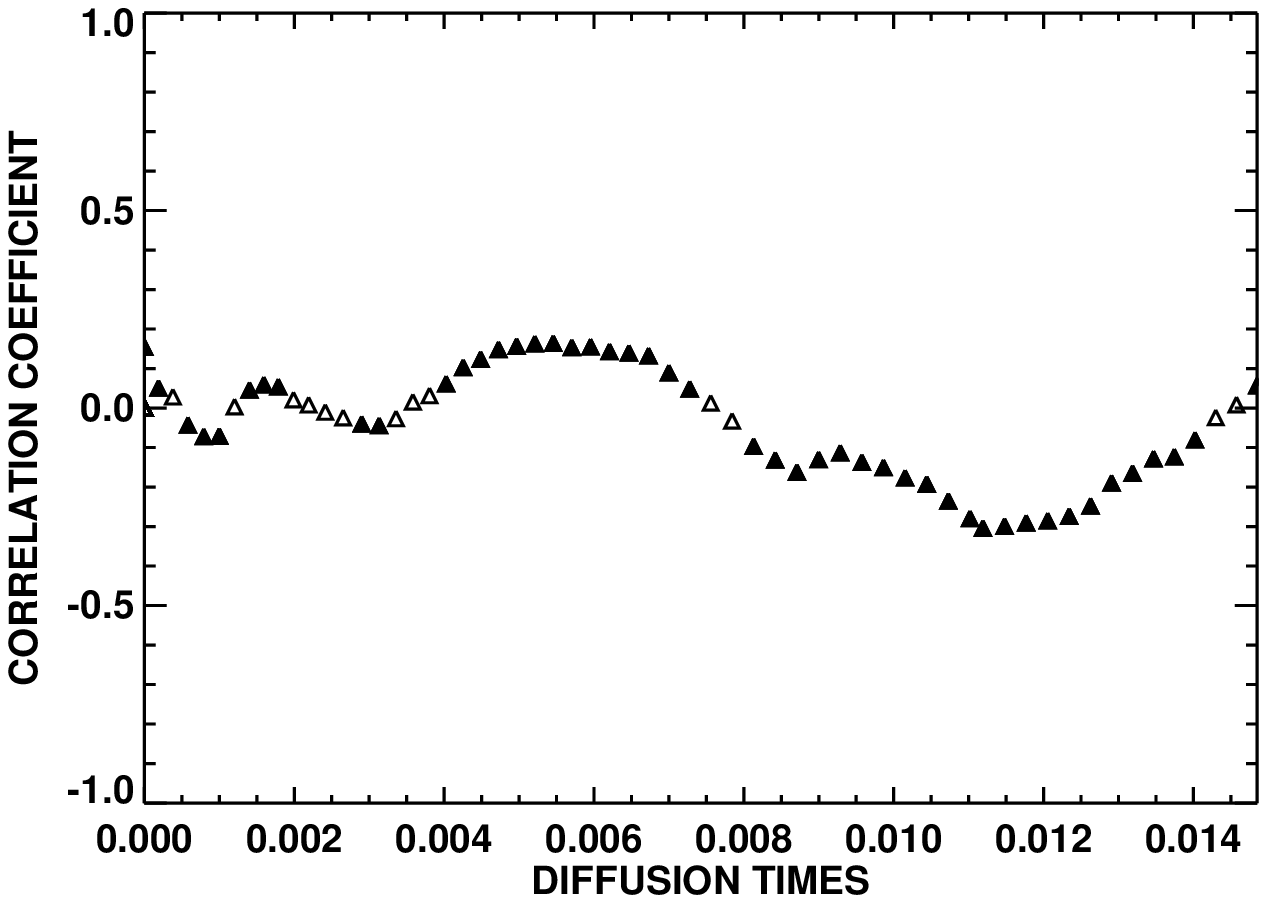}
\includegraphics[width=0.44\textwidth]{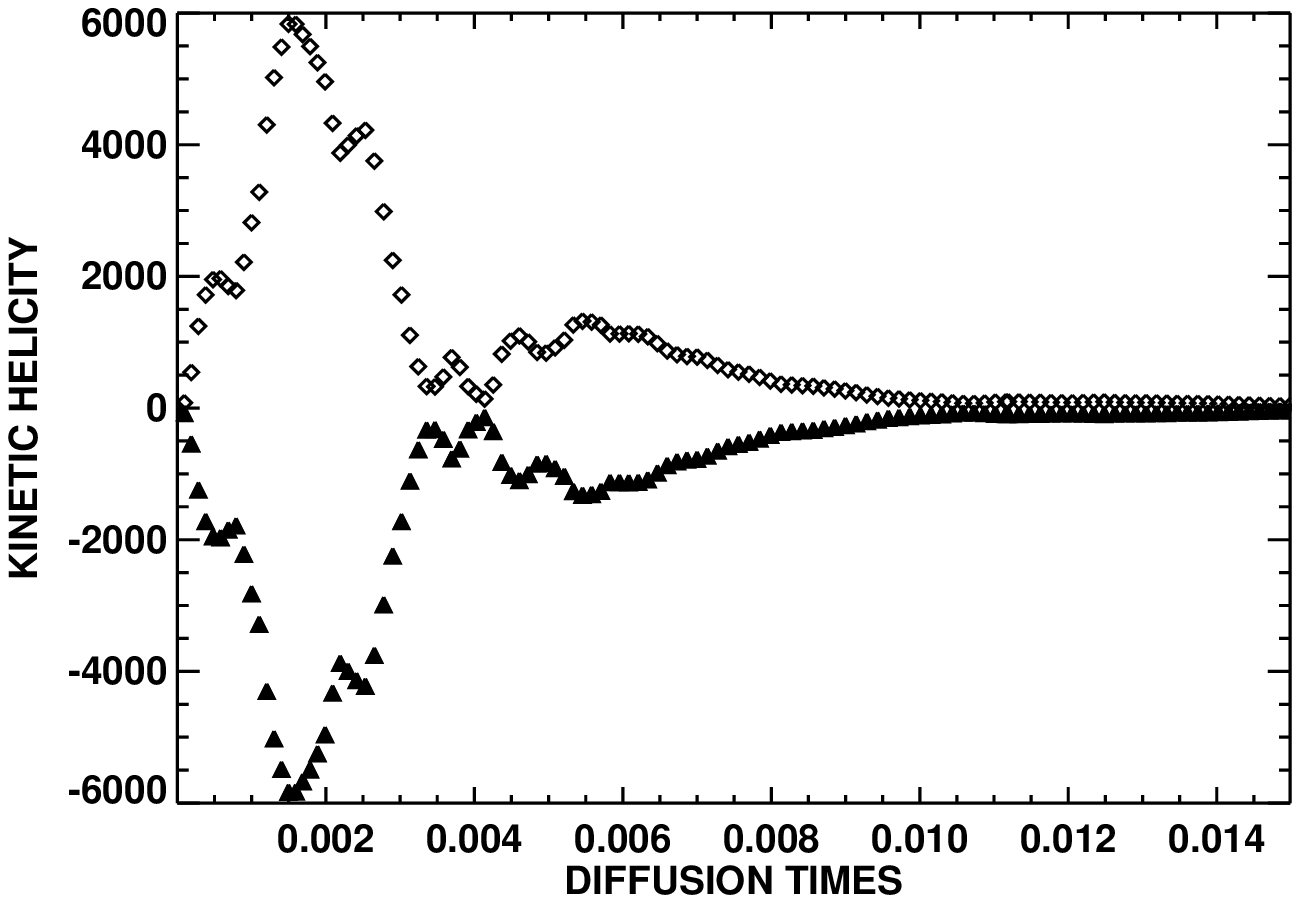}
\includegraphics[width=0.44\textwidth]{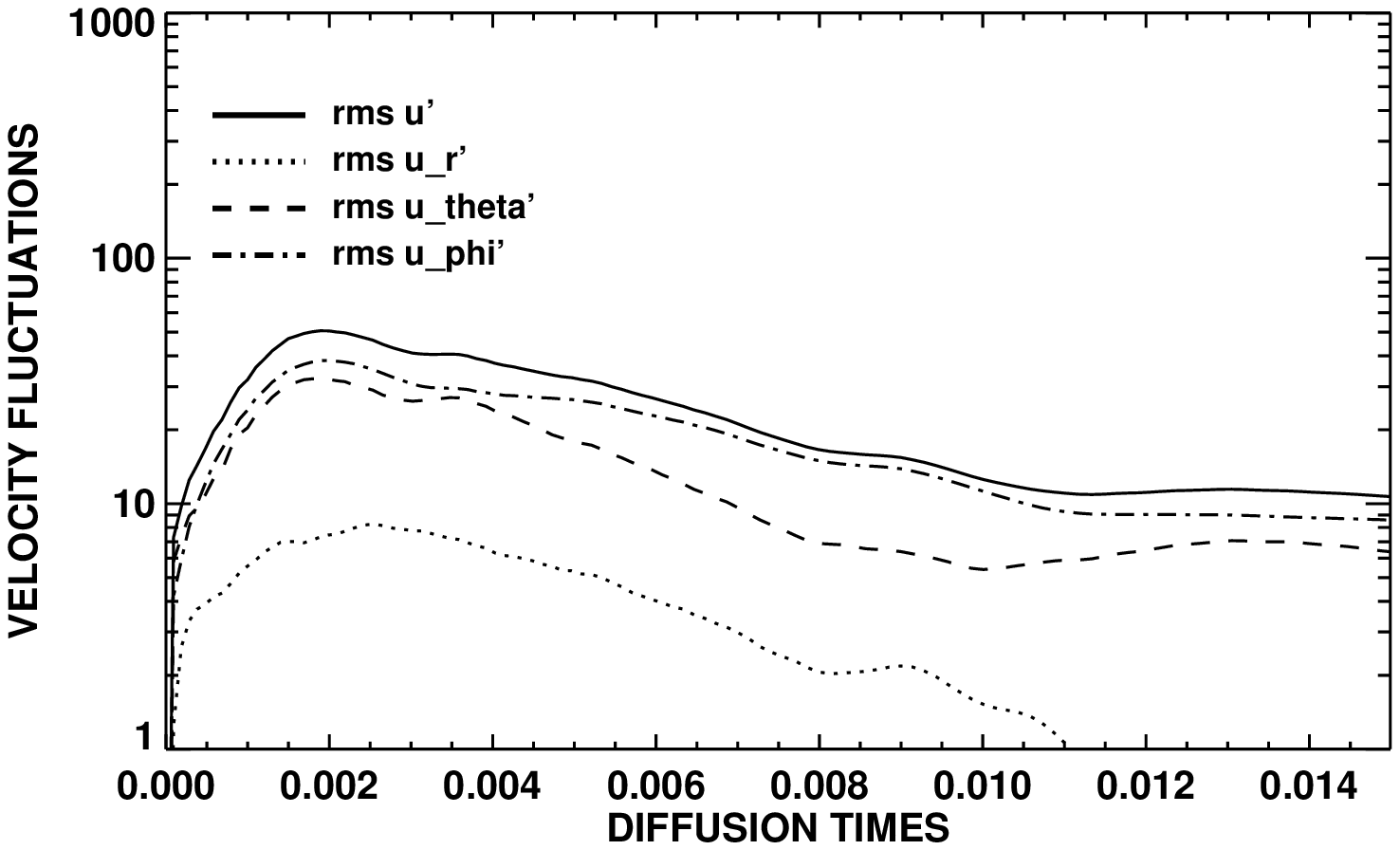}
\caption{Dynamo-effect (top-panel) expressed as $\alpha_{\phi\phi}$ measured
from the simulation with a perturbation at $t=0$. Only an initial
period of the full run of 0.015~ diffusion times length is shown
here to make the very short presence of non-zero $\alpha_{\phi\phi}$
visible. The second panel shows the correlation coefficient between
the turbulent electromotive force and $\overline B_\phi$, the third
panel shows the kinetic helicity and the bottom panel the rms values 
of the velocity fluctuations.}
\label{spuxb_nonlin11c}
\end{figure}

\begin{figure}
\includegraphics[width=0.44\textwidth]{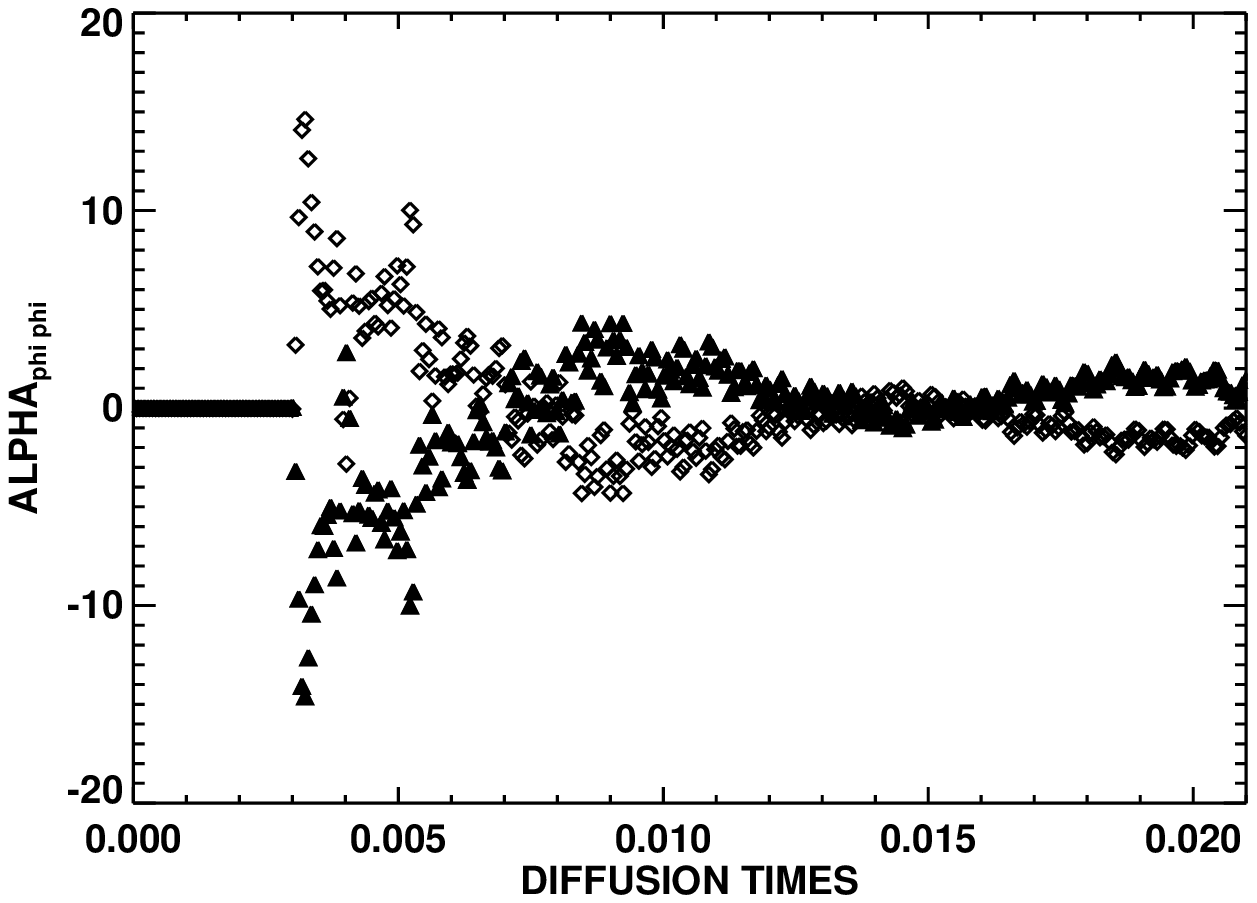}
\includegraphics[width=0.44\textwidth]{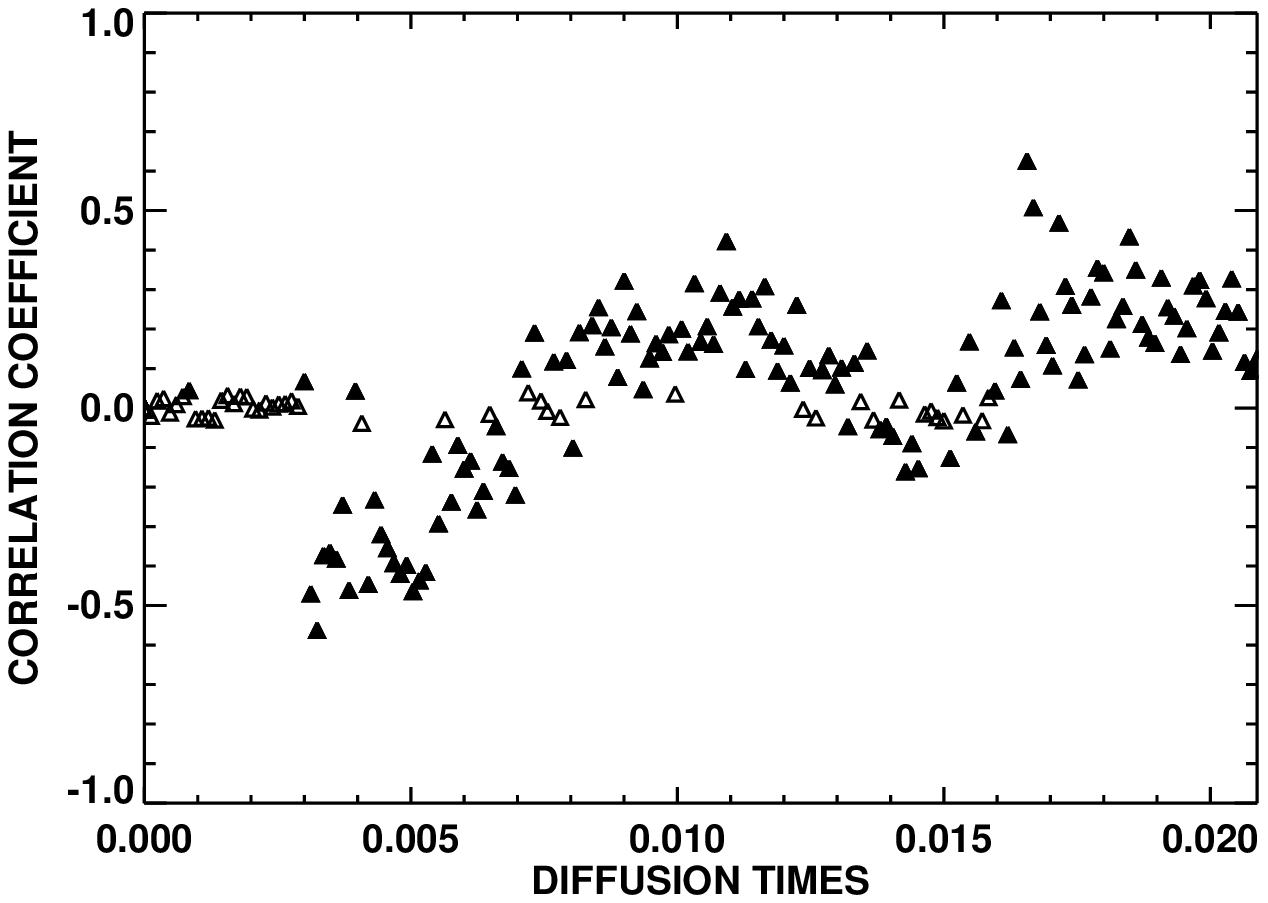}
\includegraphics[width=0.44\textwidth]{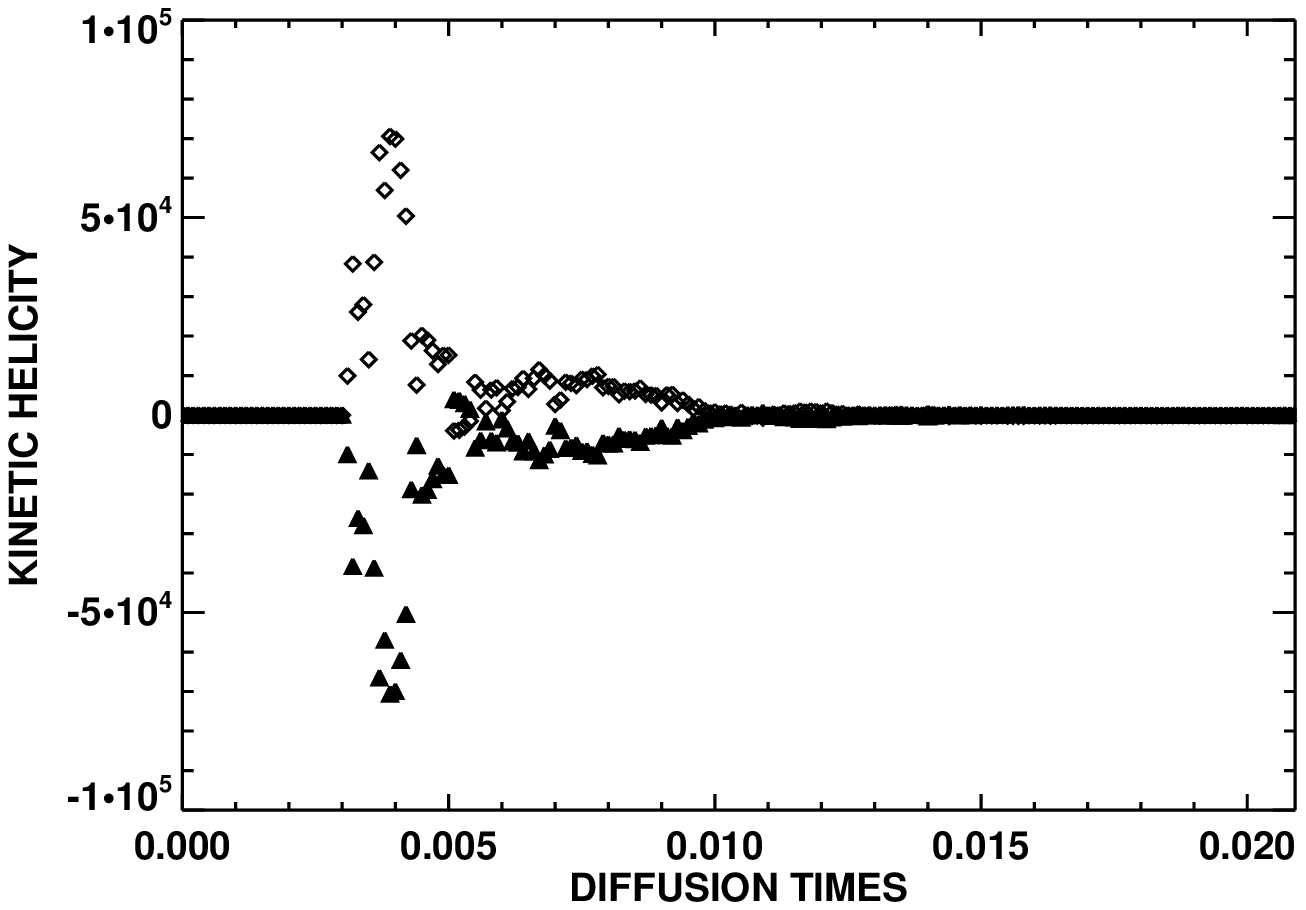}
\includegraphics[width=0.44\textwidth]{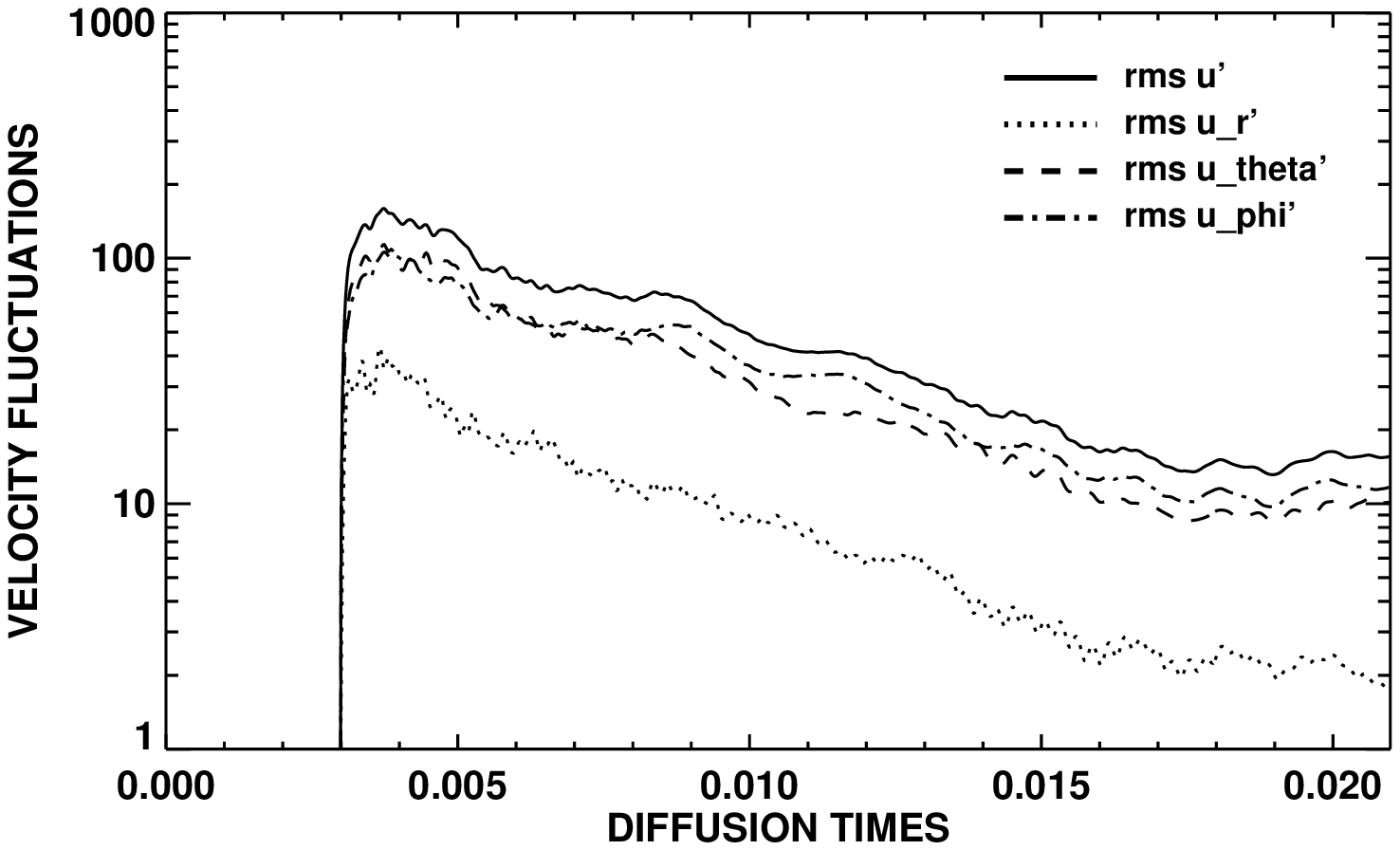}
\caption{Dynamo-effect (top-panel) expressed as $\alpha_{\phi\phi}$ measured
from the simulation with a perturbation at $t=0.003=9.5 P_{\rm rot}$. Again,
the second panel is the correlation coefficient, the middle panel shows the 
kinetic helicity and the bottom panel the rms values of the velocity fluctuations.
Note that the period covered by these plots is longer than in Fig.~\ref{spuxb_nonlin11c}
and covers about 67~rotation periods.}
\label{spuxb_nonlin16a}
\end{figure}

\begin{figure}
\includegraphics[width=0.45\textwidth]{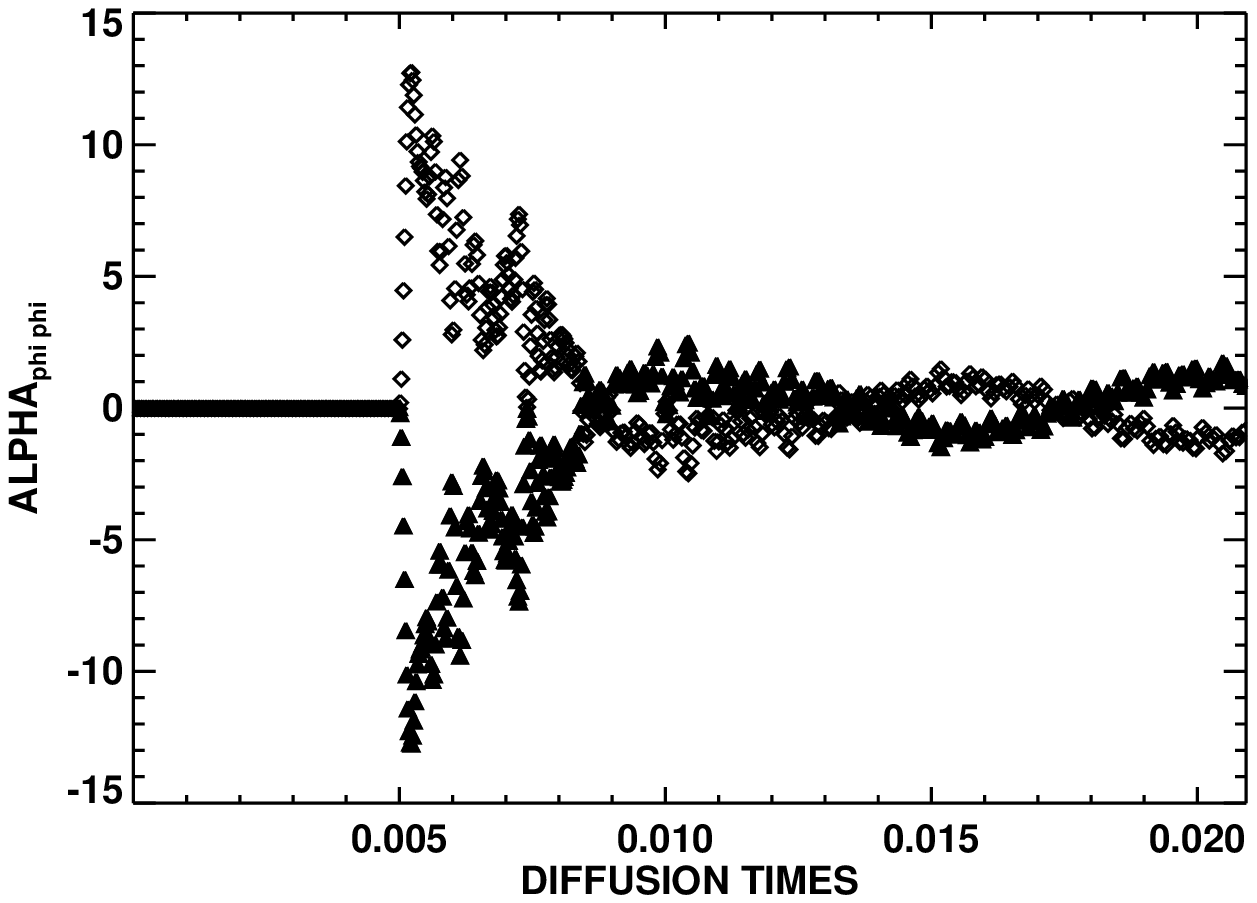}
\includegraphics[width=0.45\textwidth]{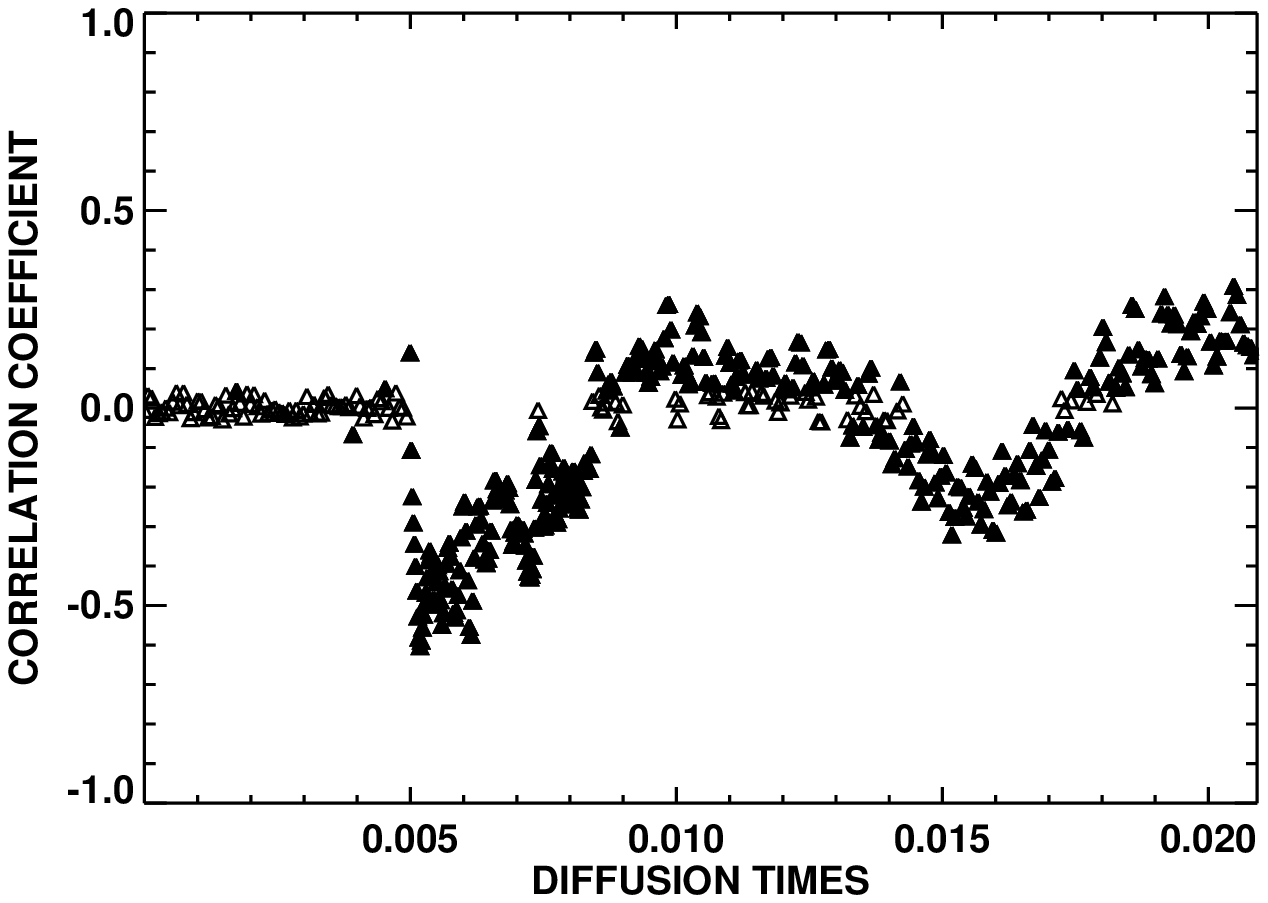}
\includegraphics[width=0.45\textwidth]{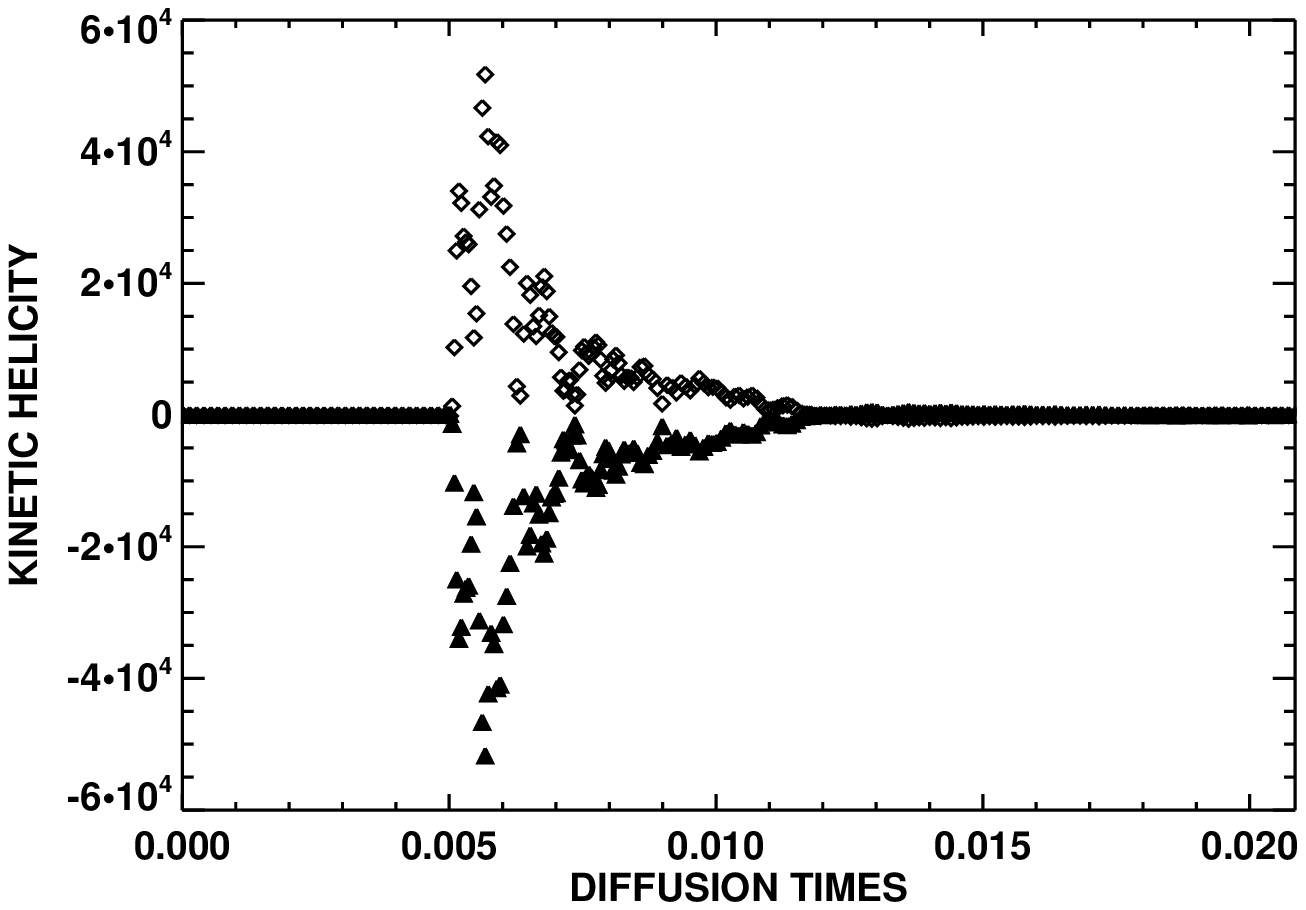}
\includegraphics[width=0.45\textwidth]{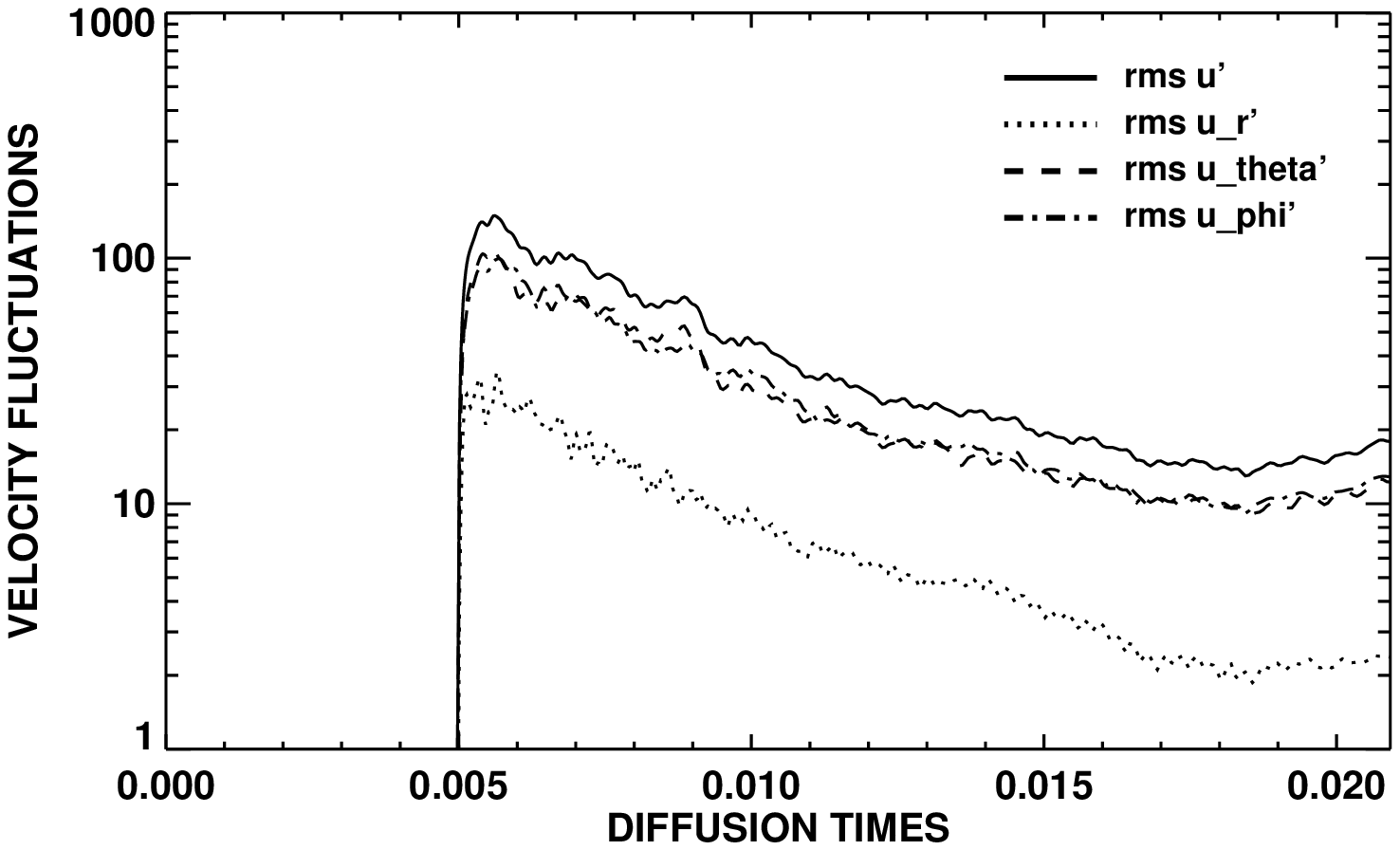}
\caption{Dynamo-effect (top-panel) expressed as $\alpha_{\phi\phi}$ measured
from the simulation with a perturbation at $t=0.005=15.9 P_{\rm rot}$. Again,
the middle panel shows the kinetic helicity and the bottom panel the rms
values of the velocity fluctuations.}
\label{spuxb_nonlin15d}
\end{figure}

Fig.~\ref{spuxb_nonlin11c} shows the estimate of $\alpha_{\phi\phi}$ 
for the run NL000 with a non-axisymmetric perturbation at $t=0$ when 
no toroidal fields were present initially. The evolution of this estimate
of $\alpha_{\phi\phi}$ is compared with measurements of the kinetic
helicity and the rms-values of the velocity components. The initial
$\alpha_{\phi\phi}$ is solely due to the initial perturbation interacting
with the system after the first time-step. The $\alpha_{\phi\phi}$
dies out very quickly -- in about one rotation period -- and oscillates 
around zero with negligible amplitude. The values of $\alpha_{\phi\phi}$
are then about 200~times smaller than the rms velocity. The correlation
coefficient of the two quantities ${\rm tEMF}_\phi$ and 
$\overline B_\phi$ in the second panel of Fig.~\ref{spuxb_nonlin11c} is 
very close to zero all the time. Open symbols are actually those cases 
for which the hypothesis of uncorrelated quantities, i.e. no $\alpha$-effect, 
holds statistically. The correlation may be still significant for the filled 
triangles. The limit was set such that there is a remaining probability of 
1~per cent for the filled symbols to represent uncorrelated quantities. 

The helicity measured compares well with the rms velocity of about 50 and
the assumption that the length-scale for the vorticity is roughly 
$r_{\rm o}-r_{\rm i}$, whence $\langle u'^{\,2}\rangle/(r_{\rm o}-r_{\rm i}) = 5000$.
The length-scale is probably slightly smaller than the thickness of the
spherical shell. We also note that the non-axisymmetric motions are
dominated by their horizontal components.

The second set of plots in Fig.~\ref{spuxb_nonlin16a} is from
model NL003 with a perturbation at $t=0.003$ which is a bit 
more than two rotation periods after the magnetic field has reached 
its Tayler-unstable strength. The values of $\alpha_{\phi\phi}$ are 
about 10~times stronger than in model NL000, and they are actually 
increasing during a short period of about one rotation. The
helicity is more than 10~times stronger than in model NL000 and
has the same sign as $\alpha_{\phi\phi}$ until $t\sim 0.0075$.
The correlation coefficient reaches values of $-0.6$.
There is apparently a second phase of a small, but significant 
$\alpha_{\phi\phi}$ which has the opposite sign than the kinetic
helicity.

The third set of plots in Fig.~\ref{spuxb_nonlin15d} shows the 
simulation NL005 with a perturbation at $t=0.005$ which is at a time
when the axi\-symmetric toroidal field has a strength of 
$\overline B_\phi=1361$ and is already decreasing, and the average
differential rotation is already near zero. At that time, the
linear stability limit is lowest in the whole period considered
here. The values of $\alpha_{\phi\phi}$ are a bit lower than those of
NL003, but are also first increasing during about one rotation period. 
Again, the correlation coefficient reaches values around $-0.6$ in this 
run. The sign change, which was also present in the $\alpha_{\phi\phi}$
of NL003, takes place at $t=0.0088$. This is relatively earlier 
than in NL003, when measured from the perturbation time.

\begin{figure}
\includegraphics[width=0.45\textwidth]{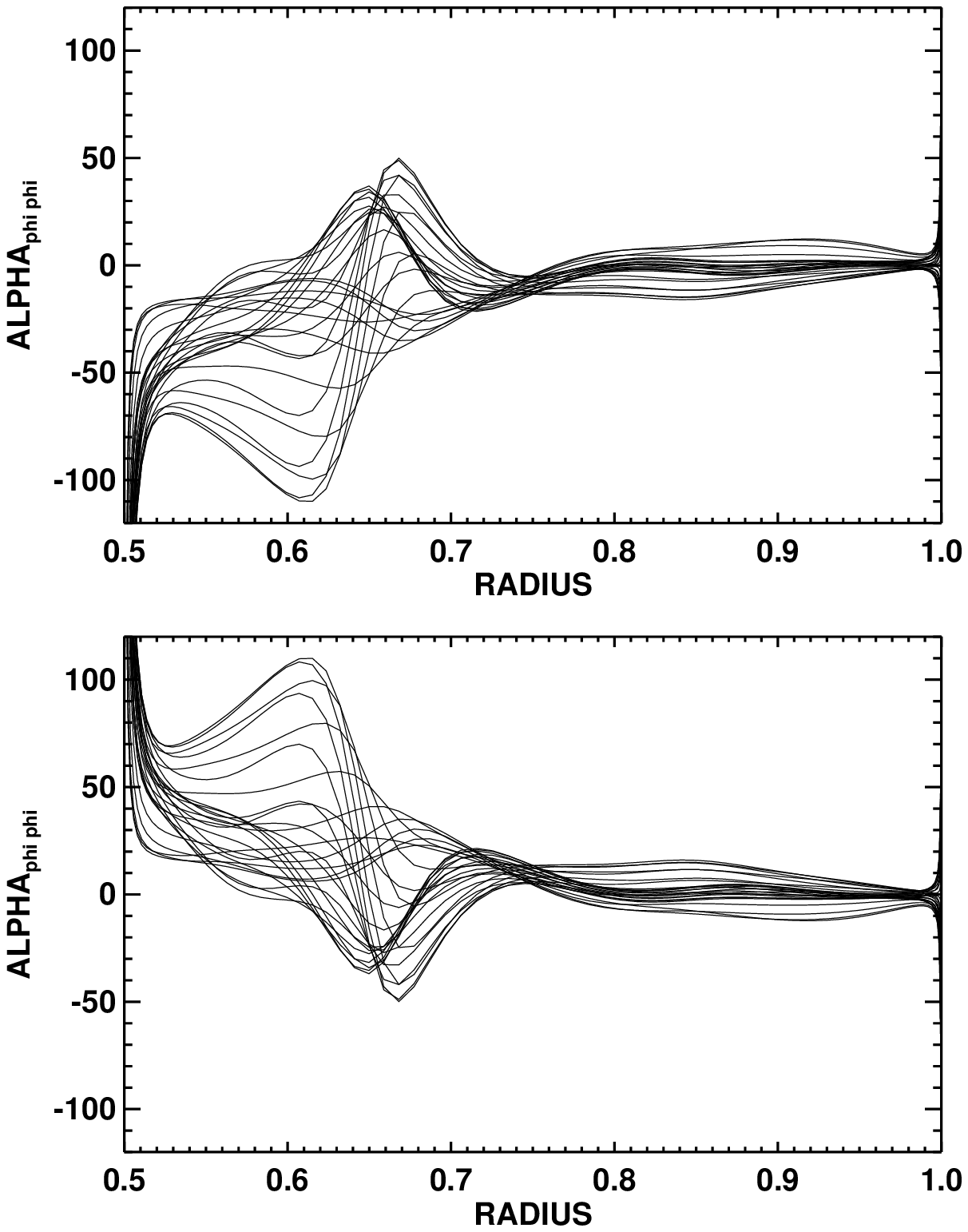}
\caption{Dynamo-effect in terms of $\alpha_{\phi\phi}$ versus
radius measured from the simulation with a perturbation at 
$t=0.003=9.5 P_{\rm rot}$. The results from 26~snapshots
with $0.0035\leq t \leq 0.004$ of the simulation are superimposed in this graph. The top
panel is for the northern hemisphere, the bottom panel is for
the southern hemisphere.}
\label{spuxb_radial16a}
\end{figure}

The radial distribution of the $\alpha$-effect can tell us something
about the origin of the dynamo action. If one uses only the values
of $\langle\vec u'\times \vec B'\rangle_\phi$ and $\overline{B_\phi}$
for a given $r$ from all the $\theta$-locations in a hemisphere, the resulting
correlation will be a local one in radius. We should bear in mind though,
that the $\alpha_{\phi\phi}$ become statistically less significant, since we 
obtain a regression line from 47~grid points only, in the cases of NL003 
and NL005. We plot the corresponding distributions of $\alpha_{\phi\phi}$ 
estimates versus radius for NL003 in Fig.~\ref{spuxb_radial16a} with 
a perturbation at $t=0.003$.

Non-zero $\alpha_{\phi\phi}$ are mostly found in the inner
half of the spherical shell, at $r<0.75$. Because of the vacuum
condition on the radial boundaries, we do not expect the turbulent 
EMF to be zero, and the $\alpha_{\phi\phi}$ derived from the
regression method thus need not vanish. A run with perfect conductor
boundary conditions at $r_{\rm i}$ was performed to evaluate the 
influence of the inner boundary. The $\alpha_{\phi\phi}$ is then indeed
zero at the inner boundary. The maximum values of $\alpha_{\phi\phi}$
near $r=0.6$ are about $\pm 90$ as compared to about $\pm 110$ in
NL003. We will show spatial distributions of $\alpha$-components again
below, obtained with another method.

Note that there is no point in looking at the total magnetic
energies in these simulations. Since neither the differential 
rotation nor the initial magnetic field are imposed anywhere
in the computational domain, there is no long-lived energy
source in the system which could drive a dynamo noticeable in
sustained magnetic energies. The concern of this Paper is
rather the evolution of stars into objects with apparently
stationary magnetic fields, such as magnetic intermediate-mass
stars. We are not focusing on sustained
dynamo action here.

\begin{figure*}
\includegraphics[width=5.6cm]{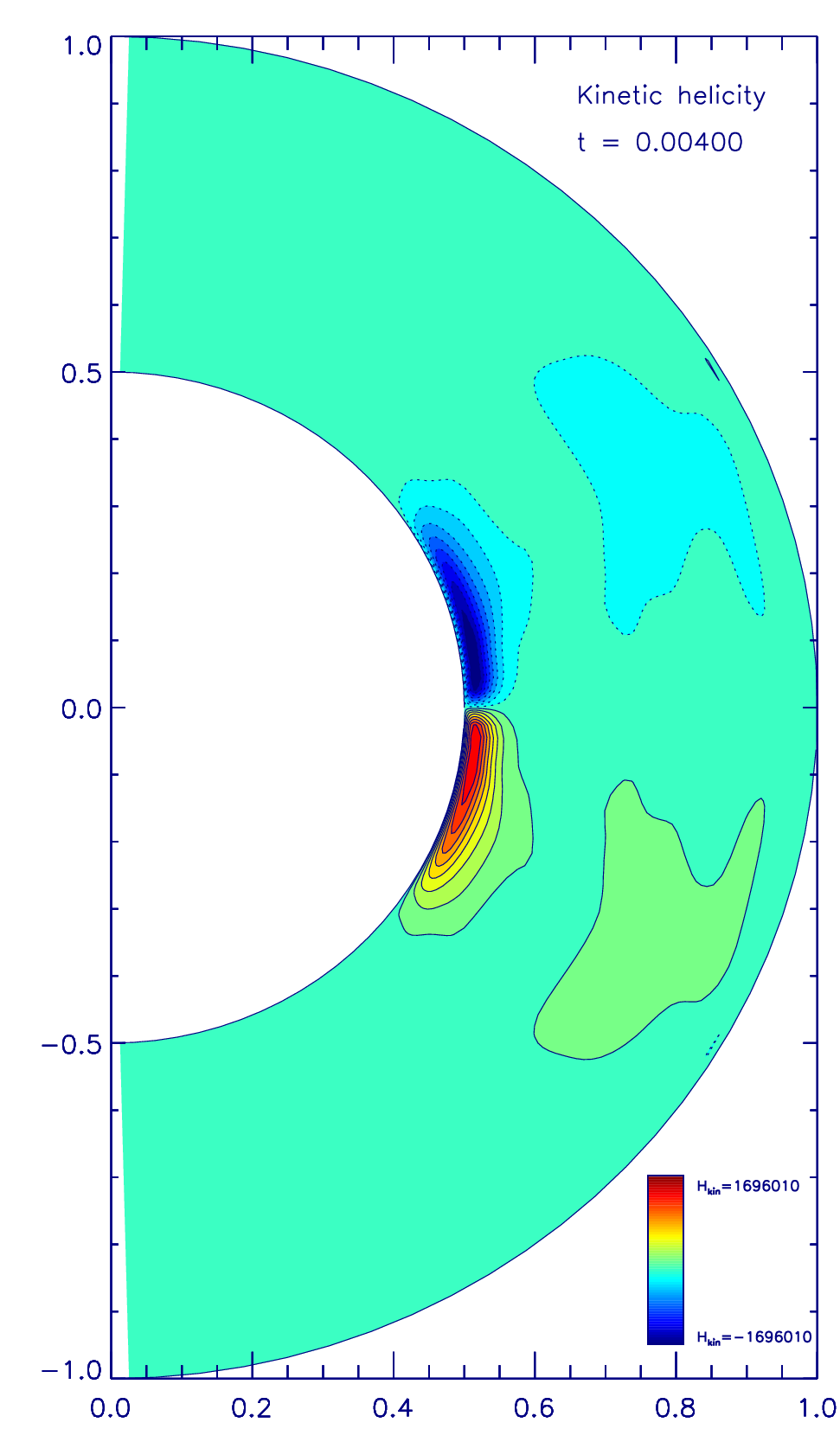}
\includegraphics[width=5.6cm]{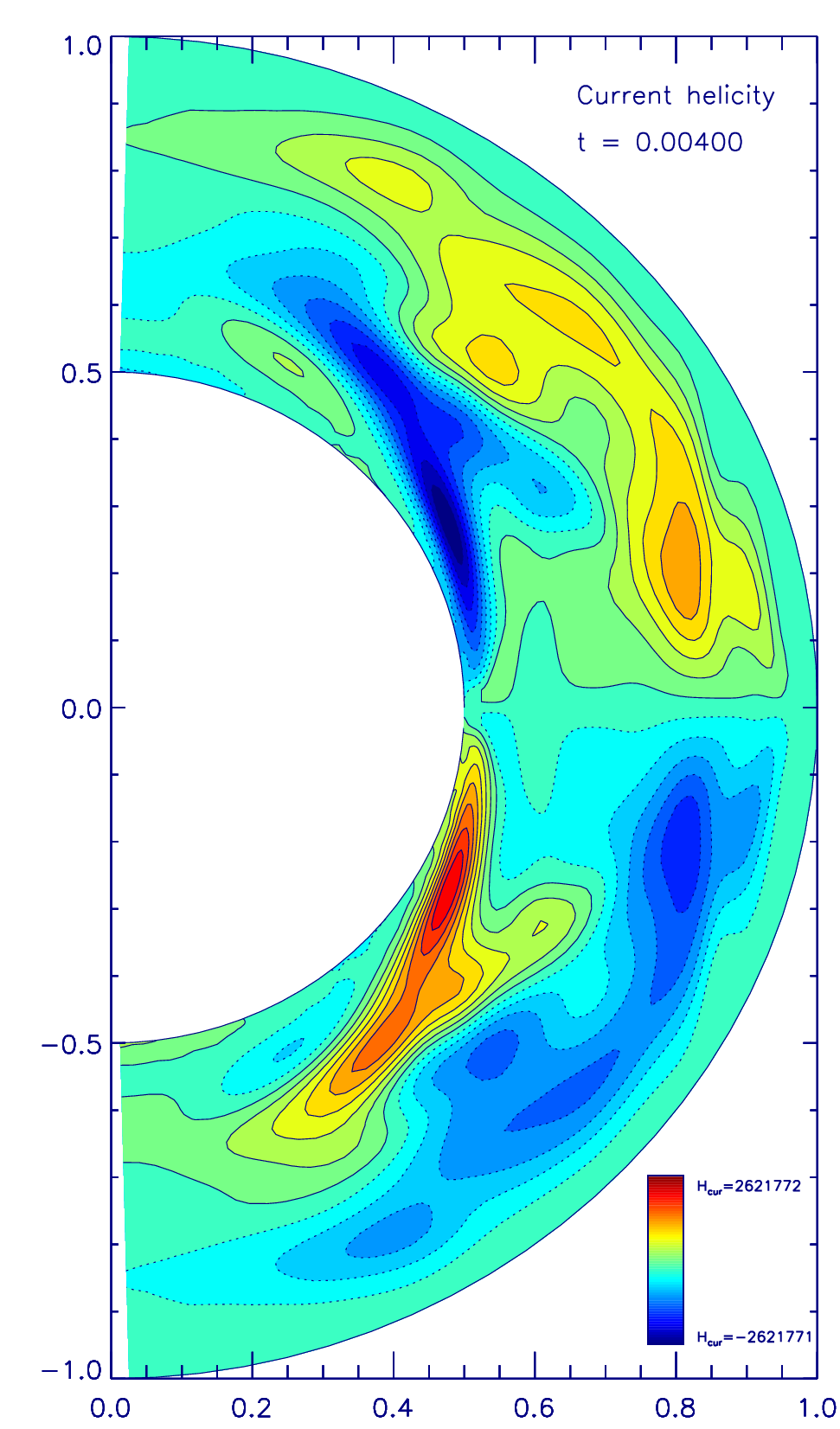}
\includegraphics[width=5.6cm]{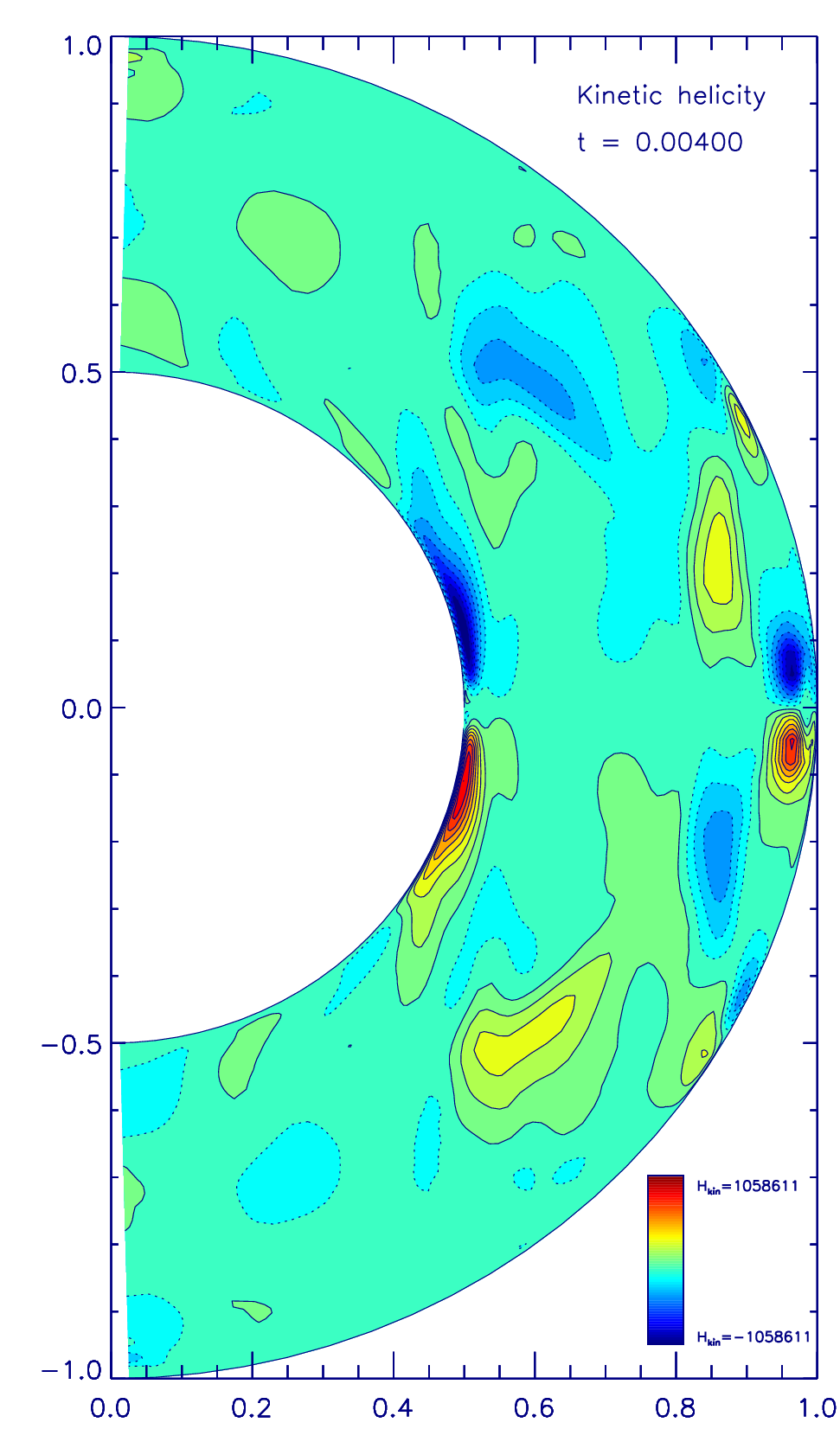}
\caption{Left: kinetic helicity in the $(r,\theta)$-plane
average over the azimuthal direction taken from the 
run NL003. Middle: current helicity obtained in the same
way for NL003. Right: kinetic helicity of the run NL003h
with doubled magnetic Reynolds number.}
\label{spshk}
\end{figure*}

The spatial distribution of the kinetic helicity and the current 
helicity is shown in Fig.~\ref{spshk}. The concentration of the 
helicity near the inner boundary resembles the result from the 
$\alpha$-measurement which also showed larger values in the inner 
part of the computational domain. More precisely, we can see that 
the helicity is actually concentrated near the place where
the tangent cylinder touches the inner sphere. 
The right panel is derived from the run NL003h which has the same
parameters as the run NL003, however, with ${\rm Rm}=40\,000$. While
the picture is more noisy than the one for lower ${\rm Rm}$, we also see
the the helicity concentration becomes thinner with respect to the
axis distance. Note that the perturbation of this run actually kicks
in at a different stage of super-critically as compared to NL003, because
rotation alters the stability limits. In general, however, we have 
to conclude that a considerable part of the kinetic helicity present 
in the system is due to the geometrical setup, namely the presence 
of an inner cylinder. The situation is unchanged in the run with 
perfect conductor boundary conditions at $r_{\rm i}$.

This is different for the current helicity. A considerable amount
of {\em positive\/} current helicity is measured in the bulk of the
northern hemisphere. We believe that it is mostly this positive
current helicity contributing to the small, but positive 
$\alpha_{\phi\phi}$-effect in the bulk of the northern hemisphere.
The regression method is a bit crude at this point; the below results 
from the test-field method will show this more clearly.

It is interesting to note that the sign of $\alpha_{\phi\phi}$ and 
the kinetic helicity is the same in the inner half of 
the radial extent and especially near the tangent cylinder. This
contradiction with (\ref{helicity}) is apparently caused by a
one-cell helical motion, rather than by an average of turbulent
motions. Interestingly, Reshetnyak (2006) also finds negative
kinetic helicity along the tangent cylinder, albeit it is a
convective, non-magnetic simulation. The feature is more obvious
in the run with low Rayleigh number, in which convection is not
reaching a turbulent state. We think that the negative helicity
near the tangent cylinder of the northern hemisphere is an
inner-boundary effect, neither related to convective nor Tayler
instability turbulence.


\begin{figure*}
\includegraphics[width=0.8\linewidth]{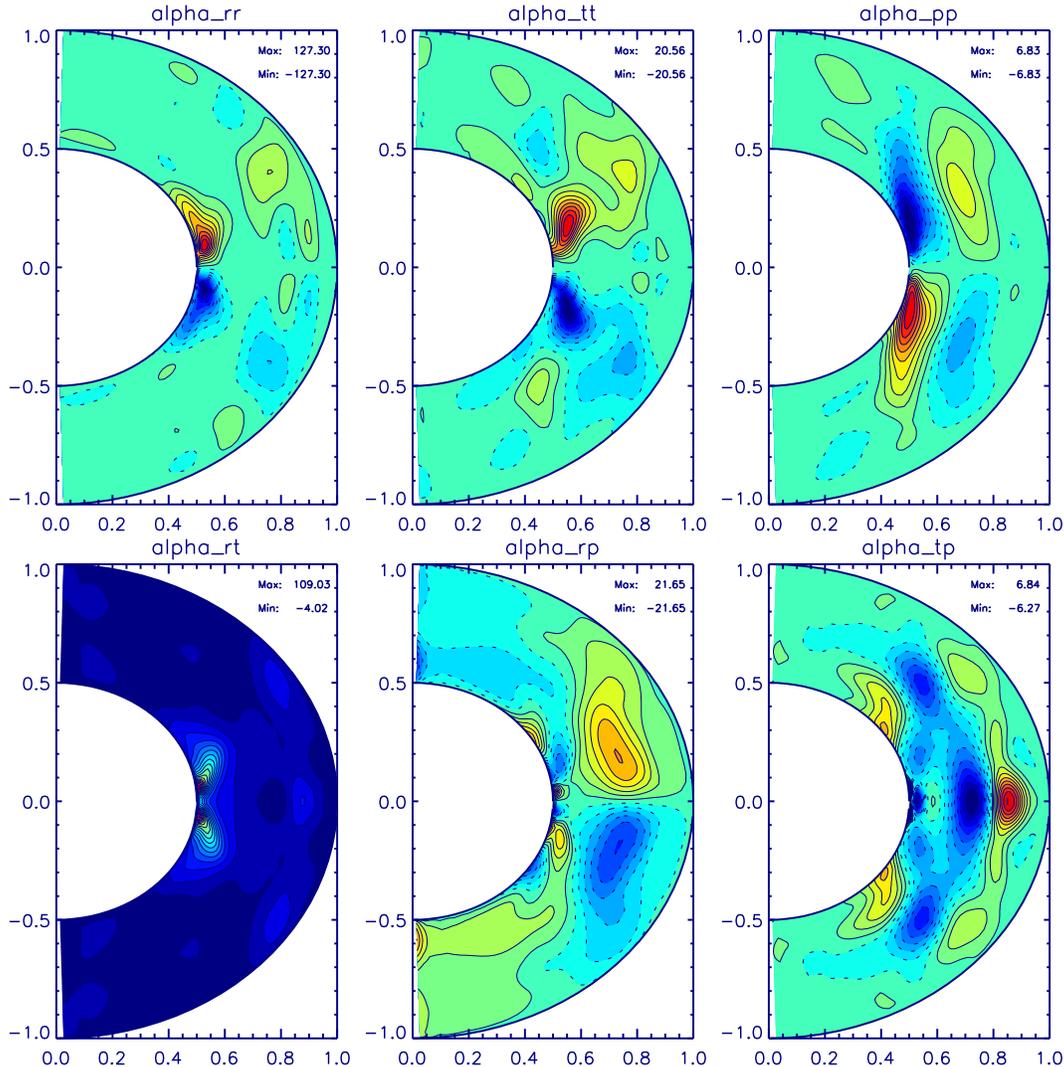}
\caption{Coefficients of the symmetric $\alpha$-tensor obtained by the
test-field method. Units are in the dimensionless units of velocity of
the simulation. The cross-sections shown here are averaged within 
$t=0.0035$ and $t=0.0040$ diffusion times.}
\label{spdyn_nonlin16a_176_200_alpha}
\end{figure*}

\subsection{Mean-field coefficients from the test-field method}
The second attempt to look for dynamo action consists of the
test-field method as described by Schrinner et al. (2007).
In parallel to the actual non-linear simulation, several
equations for a magnetic-field like quantity $\vec b^{(i)}$ 
similar to the induction equation are evolved,
following the action of the large-scale and small-scale velocity,
$\overline{\vec u}$ and $\vec u'$, from the real simulation on 
given test-fields $\overline{\vec B}_{\rm T}^{(i)}$. Because of the similarity of the computational
setup use here and of the simulations used by Schrinner et al. (2007),
we implemented the test-field method precisely in the same way.
Note the different definition of the signs of $\vec\alpha$ and
$\vec\gamma$ in (\ref{emf}) though.

The separation of large-scale and small-scale contributions to 
$\vec u$ and is again done by azimuthal averages, leading to 
a large-scale field depending on $r$ and $\theta$ only. Nine
different test fields give rise to 27~test-field equations,
delivering nine mean electromotive forces $\vec{\cal E}^{(i)}$ from $\vec u'$ and
the $\vec b^{(i)}$. Replacing $\vec{\rm EMF}$ and $\overline{\vec B}$
by $\vec{\cal E}^{(i)}$ and $\overline{\vec B}_{\rm T}^{(i)}$
in (\ref{emf}) leads to a system of equations which can
be solved for the mean-field coefficients.

The values of the symmetric $\vec\alpha$-tensor averaged over time are shown
in Fig.~\ref{spdyn_nonlin16a_176_200_alpha}, taken from the run
NL003 and the period from $t=0.0035$ to $t=0.0040$. The initial 
phase of the onset of the instability has thus been omitted. We
see that it is $\alpha_{rr}$ which has the strongest peak values,
followed by $\alpha_{\theta\theta}$ and $\alpha_{\phi\phi}$. In
an axisymmetric mean-field dynamo, $\alpha_{rr}$  and $\alpha_{\theta\theta}$ 
are generators of the toroidal magnetic field. While $\alpha_{\phi\phi}$ 
has the same sign as the kinetic helicity near the inner boundary, the 
other two diagonal elements have the opposite sign. The anisotropy
of the $\alpha$-effect is due to the relatively strong rotation of
the system. The numbers can be directly compared to the velocity
measurements shown in Fig.~\ref{spuxb_nonlin15d}.

The contributions from the derivatives of $\overline{\vec B}$ are
rather small. The coefficients $\beta_{rr}$, $\beta_{\theta\theta}$,
and $\beta_{\phi\phi}$ are all positive and do not exceed the order
or unity. Interestingly, the contributions from several pairs of
coefficients nearly cancel in the construction of the turbulent
EMF. In particular, these are $\beta_{r\phi}$ cancelling with $-\delta_\theta$, 
and $\beta_{\theta\phi}$ with $\delta_r$.

\subsection{Comparison}
Both methods have their disadvantages. On the one hand, the regression 
method for the determination of $\alpha_{\phi\phi}$ ignores the possible
influence of other mean-field coefficients on the $\phi$-component of 
the turbulent EMF. The test-field method, on the other hand, relies on 
turbulence which is not magnetically driven. Only the velocity field 
from the actual simulation enters the determination of the mean-field 
coefficients. Our simulations are concerned with magnetically
generated fluctuations $\vec u'$ and $\vec B'$, however. It
is therefore very likely that the mean-field coefficients determined
by the test-field method are systematically underestimated. A
comparison of the rms magnetic fields, $B_{\rm rms}=\sqrt{\langle B'^2\rangle}$, with 
rms velocities, $u_{\rm rms}=\sqrt{\langle u'^2\rangle}$, over time
reveals a factor of about three between the two for the run NL003.

It is elucidating to compare the turbulent EMF measured from 
the simulation directly with the turbulent EMF constructed from 
the mean-field coefficients together with the measured large-scale 
$\overline{\vec B}$. The directly measured EMF is about four 
times larger than the reconstructed EMF. This ratio is very similar 
to the above mentioned ratio of $B_{\rm rms}$ to $u_{\rm rms}$. The
current helicity shown in Fig.~\ref{spshk} in the shell from
$r=0.75$ to $r=0.95$ is even ten times larger than the kinetic 
helicity in the same region.


\section{Summary and discussion}
We studied the amplification and stability of
magnetic fields in the pre-main-sequence evolution of intermediate-mass
stars. Especially the phase when the stars possess an extensive radiative 
envelope already is considered. The amplification of a poloidal magnetic 
field by differential rotation is studied in a non-convective, spherical 
shell. The feedback by the Lorentz force diminishes the differential
rotation brings the field growth to a halt. The amplification time
and amplitude only depend on the initial, poloidal magnetic field. The
amplification time is given by 
\begin{equation}
  \Delta t \approx R\sqrt{\mu\rho}/B_r
\end{equation}
and results to about 7500~yr for a star with $R=3{\rm R}_\odot$, 
$\rho = 0.1$~g\,cm$^{-3}$, and an initial magnetic field
strength of $B_r = 1$~G. These time-scales are very short, but
one needs to bear in mind that the stellar winds in this phase
of the evolution will still remove angular momentum and partly
sustain the internal differential rotation. It is very likely
that the whole process of amplification and onset of the instability
take much longer in reality than suggested by our initial-value
simulations.

Instability occurs fairly early at an Alvf\'en angular velocity
of about $\Omega_{\rm A} = RB_\phi = 1000$. This is much smaller
than the condition for stability, $\Omega^2 < \Omega_{\rm A}^2$,
derived by, e.g., Pitts \& Tayler (1985). The difference is the
much more complex structure of $B_\phi$ in our case with much
stronger currents than the configuration used for the analytical
study, where $B_\phi = s$, $B_z = {\rm const}$. The fraction
of the critical Alfv\'en angular velocity to the actual rotation,
$\Omega_{\rm A}/\Omega$ was even smaller in the stability analysis 
of the solar tachocline by Arlt, Sule \& R\"udiger (2007) where 
the toroidal magnetic fields were even more localised.

Test runs on the linear stability (step~II) indicate the a positive
shear has a stronger stabilizing effect than negative shear in
our particular configuration. The test was performed on the toroidal
field at $t=0.002$ from step~I with $B_0=300$ and an arbitrary
angular velocity $\Omega(s)$ increasing by 15 per cent from $s=0$ to
$s=1$. The stability limit was at $S>2$, while the stability limit
for the same non-axisymmetric mode with $d\Omega/ds<0$ was $S=1.2$. 
This explains the steep increase
of the lines of marginal stability in Fig.~\ref{stability} after
$t\sim 0.06$. In the pre-main-sequence evolution, it may be
responsible for a long-term of stabilization as long as the star
is gaining angular momentum on the surface by accretion. This
situation is also stable against axisymmetric perturbations
exciting the magnetorotational instability. It will be very 
interesting to combine the simulations in this paper with the 
actual angular-momentum evolution obtained from combining the 
effects of accretion, magnetic star-disk coupling, and magnetized 
winds.

During most of the pre-main-sequence period, the star is rotating
rather rapidly, with rotation periods of 1--2~days. Rapid rotation
will suppress the Tayler instability in general. Once the
accretion ceases and the disk turns into a more passive
environment of the star, the stellar wind leads to a
spin-down of the star. The increase of the rotation period
can be quite steep, but only if strong magnetic fields are
present (St\c{e}pie\'n \& Landstreet 2002). We are left
with a contradiction here, since the spin-down requires
poloidal fields, but the Tayler-instability providing
these poloidal fields will set in only if the rotation
period has already increased.

Another question concerns the time it takes for the unstable
magnetic field to become visible on the surface. In the
3D simulation, we find a maximum of $B_{\rm max}^{\rm (surf)}=168$ at the 
surface, about 18~rotation periods after the onset of the
instability. Note that the initial poloidal magnetic field
was internal; the surface field was zero, but it develops
surface fields of the order of $B_{\rm max}^{\rm (surf)}\sim 10$
by diffusion. The bottom panel in Fig.~\ref{spuxb_nonlin16a}
demonstrates that the motions arising from the instability
are mostly horizontal. The emergence of the flux at the
stellar surface is therefore rather a diffusive process.
In the numerical simulation, the dynamical and diffusive
time-scales are not as widely separated as in reality. We
may argues that the emergence time is a diffusive one.
Given a length-scale of say $0.1{\rm R}_\odot$ for the rise
and a microscopic diffusivity of $1000~{\rm cm}^2~{\rm s}^{-1}$, 
this time-scale results in about 9~Myr. The true rise time is 
certainly a combination of several effects and may be faster. 
The rise of a {\em stable\/} toroidal flux tube was studied by 
Mestel \& Moss (2010). A tentative conversion of their
time-scales into stellar values leads to 150~Myr and more
which is more than an order of magnitude longer than in 
the unstable case (smaller structures).

The maximum poloidal field at the surface converts to about 
2.7~kG using (\ref{bphys}). The time-scale of 9~Myr is 
actually very interesting for Ap stars, it may take some 
time on the main sequence before the Ap phenomenon is 
observable (Hubrig, North \& Mathys 2000). Other observations
favour a persistence of magnetic fields from early Herbig Ae/Be
stars to main-sequence Ap stars (Wade et al. 2007). By contrast,
Hubrig et al. (2009) find decreasing magnetic fields in Herbig
Ae/Be stars with age. While the observational picture will be 
gradually completing, simulations for a whole 
sequence of magnetic Reynolds numbers will be need 
to tell whether the emergence time-scale is indeed a fraction 
of the diffusion time. Note that the above mentioned 2.7~kG are
small compared to the internally possible toroidal fields of 
$10^5$--$10^6$~G. What appears to be a strong field in observations,
can yet be a remnant of something much larger. 

Also the amount of complexity of the
surface magnetic fields is compatible with surface fields
from Zeeman Doppler imaging (e.g. Kochukhov et al. 2004).
Note that differences in initial conditions may cause a 
variety of final surface field strengths as well as a variety 
of emergence times, very similar to the complex observational
picture that has been compiled up to now.

We are left with two possible ways to magnetic Ap stars:
the existence of fossil field configurations that are
stable for very long times on the one hand, and the emergence 
of magnetic fields by an instability on the other hand, where 
the `disrupted' configuration is the observed and long-term 
stable one. Our paper studies the latter option. Since the 
differential rotation vanishes due to the presence of the 
fields, there is no further build-up of toroidal fields and 
the `disrupted' fields will be fairly stationary over evolutionary 
time-scales. A comparison of the topology at very late stages
with the ones obtained in compressible, but non-rotating simulations
by Braithwaite \& Nordlund (2006) will be interesting, once
long enough simulations are available.

It is probable that the pre-main-sequence evolution of
A stars is a mixture of various phenomena: the presence
or absence of a substantial magnetic field in the collapsing 
cloud core, a possible short-lived dynamo during the
convective phase (Arlt 2009), the angular-momentum evolution controlled
by various torques, and the onset of the Tayler instability.
It is still a challenge to explain why roughly 10~per cent  of the
possible scenarios lead to the Ap star phenomena, while
the remaining fraction does not lead to strong fields at
the stellar surfaces (fields may still be hidden in the star).
However, the study of differences in the rotational evolution 
including differences in the amplitude of the differential 
rotation (delivering both a wide range of toroidal fields and
large difference in stability limits) are a promising field 
to find discriminating situations between normal A stars and 
Ap stars.

{\em Dynamo action\/} -- Apart from looking at the the evolution of magnetic fields in 
the pre-main-sequence phase of Ap stars, we also studied the 
possibility of driving a dynamo from the instability in terms 
of representing it as a mean-field dynamo.
We find enhanced values for the dynamo-$\alpha$ and related
effects describing the field generation in a mean-field context.
An axisymmetric dynamo driven by the Tayler instability cannot
be excluded, although a proper energy source is missing in our
setup. The prerequisite for the existence of an $\alpha$-effect
appears to be the presence of an inner boundary at which most of 
the $\alpha$ is concentrated, as is the kinetic helicity. The
situation is thus interesting for Earth-like planetary interiors
which have a solid inner core acting as an inner wall. To be 
interesting for a radiative zone, an inner boundary will be 
required such as a convective core, whose turbulent viscosity 
has a similar effect like a wall. This core should not be too
far away from the Tayler-unstable zone, so a possible dynamo
could be relevant for early B stars at best. The results do not
support a dynamo from the Tayler instability in the solar 
tachocline though.

An interesting domain is the Earth's dynamo though, since the
outer, fluid spherical shell of the core is very similar to what
was studied here. Because of the rather small differential rotation 
of the core, the onset of current-driven instabilities as well as 
helicity distributions as shown in this Paper are possibly emerging 
in geodynamo simulations.

\section*{Acknowledgments}
The authors are grateful for the hospitality of NORDITA, Stockholm, where the 
idea of applying the test-field method to the Tayler problem materialized. 
We would also like to thank Swetlana Hubrig for helpful discussions.

\bsp

\label{lastpage}


\begin{thebibliography}{99}
\bibitem{10}
Arlt R., Sule A., R\"udiger G., 2007, A\&A, 461, 295
\bibitem{12}
Arlt R., 2009, in Strassmeier K.~G., Kosovichev A.~G., Beckman J.~E., eds., 
Proc. IAU Symp 259: Cosmic Magnetic Fields: From Planets, to Stars and Galaxies. 
CUP, Cambridge, p.~443
\bibitem{15}
Braithwaite J., 2006a, A\&A, 449, 451
\bibitem{20}
Braithwaite J., 2006b, A\&A, 453, 687
\bibitem{30}
Braithwaite J., Nordlund \AA., 2006, A\&A, 450, 1077
\bibitem{40}
Brun A.~S., Zahn J.-P., 2006, A\&A, 457, 665
\bibitem{64}
Dikpati M., Cally P.S., Gilman P.A., 2004, ApJ, 610, 597
\bibitem{70}
Gellert M., R\"udiger G., Elstner D., 2008, A\&A, 479, L33
\bibitem{72}
Giesecke A., Ziegler U., R\"udiger G., 2005, Phys. Earth Planet. Int., 152, 90
\bibitem{74}
Gilman P.~A., Fox P.~A., 1997, ApJ, 484, 439
\bibitem{75}
Gough D.~O., McIntyre M.~E., 1998, Nature, 394, 755
\bibitem{76}
Hollerbach R., 2000, Int. J. Num. Meth. Fluids, 32, 773
\bibitem{78}
Hubrig S., North P., Mathys G., 2000, ApJ, 539, 352
\bibitem{83}
Hubrig S., et al., 2009, A\&A, 502, 283
\bibitem{87}
K\"apyl\"a P.~J., Korpi M.~J., Brandenburg A., 2009, A\&A, 500, 633
\bibitem{88}
K\"uker M., R\"udiger G., 2008, J. Phys. Conf. Ser. 118, 012029
\bibitem{89}
Kochukhov O., Bagnulo S., Wade G.~A., Sangalli L., Piskunov N., Landstreet J.~D.,
Petit P., Sigut T.~A.~A., 2004, A\&A, 414, 613
\bibitem{90}
Mestel L, Moss D., 2010, MNRAS, 405, 1845
\bibitem{91}
Mikula\v{s}ek Z., et al., 2008, A\&A, 485, 585
\bibitem{93}
Pitts E., Tayler R.~J., 1985, MNRAS, 216, 139
\bibitem{94}
Reshetnyak M.~Yu., 2006, Izvestiya, Phys. Solid Earth, 42, 449
\bibitem{96}
R\"udiger G., Kitchatinov L.~L., 1997, Astr. Nachr., 318, 273
\bibitem{98}
R\"udiger G., Kitchatinov L.~L., 2010, Geophys. Astrophys. Fluid Dyn., 104, 273
\bibitem{99}
Schrinner M., R\"adler K.-H., Schmitt D., Rheinhardt M., Christensen U.~R., 2007,
Geophys. Astrophys. Fluid Dyn., 101, 81
\bibitem{100}
Spruit H., 1999, A\&A, 349, 189
\bibitem{102}
Spruit H., 2002, A\&A, 381, 923
\bibitem{104}
St\c{e}pie\'n K., 2000, A\&A, 353, 227
\bibitem{105}
St\c{e}pie\'n K., Landstreet J.~D., 2002, A\&A, 384, 554
\bibitem{110}
Tayler R.~J., 1973, MNRAS, 161, 365
\bibitem{120}
Vandakurov Yu.~V., 1972, SvA, 16, 265
\bibitem{127}
Wade G.~A., Bagnulo S., Drouin D., Landstreet J.~D., Monin D., 2007, MNRAS, 376, 1145
\bibitem{130}
Watson M., 1981, Geophys. Astrophys. Fluid Dyn., 16, 285
\bibitem{140}
Zahn J.-P., Brun A.~S., Mathis S., 2007, A\&A, 474, 145
\end{thebibliography}
\end{document}